\documentclass{article}
\usepackage{amsmath,amssymb,amsfonts}
\usepackage{graphicx}
\usepackage{subcaption}
\graphicspath{ {/plots/without_edge_filtering/} }
\usepackage{url}
\usepackage{xcolor}
\usepackage{dirtytalk}
\usepackage{todonotes}
\usepackage{pdfcomment}
\pdfcommentsetup{draft} 
\pdfcommentsetup{color={1.0 1.0 0.0}, open=true}

\author{Alexander Shevtsov \\   \href{shevtsov@ics.forth.gr}{shevtsov@ics.forth.gr} 
   \and Maria Oikonomidou \\   \href{mailto:mareco@ics.forth.gr}{mareco@ics.forth.gr}
   \and Despoina Antonakaki \\   \href{mailto:despoina@ics.forth.gr}{despoina@ics.forth.gr}
   \and Polyvios Pratikakis \\   \href{mailto:polyvios@ics.forth.gr}{polyvios@ics.forth.gr}
   \and Alexandros Kanterakis \\ \href{mailto:kantale@ics.forth.gr }{kantale@ics.forth.gr}
   \and Sotiris Ioannidis \\   \href{mailto:sotiris@ics.forth.gr}{sotiris@ics.forth.gr}
   \and Paraskevi Fragopoulou \\   \href{mailto:fragopou@ics.forth.gr}{fragopou@ics.forth.gr} }

\begin{document}

\title{Discovery and classification of Twitter bots}

\maketitle

\begin{abstract}
A very large number of people use Online Social Networks daily.
Such platforms thus become attractive targets for agents that seek
to gain access to the attention of large audiences, and influence
perceptions or opinions.  Botnets, collections of automated accounts
controlled by a single agent, are a common mechanism for exerting
maximum influence.  Botnets may be used to better infiltrate the
social graph over time and to create an illusion of community
behavior, amplifying their message and increasing persuasion.

This paper investigates Twitter botnets, their behavior, their
interaction with user communities and their evolution over time.
We analyzed a dense crawl of a subset of Twitter traffic, amounting to
nearly all interactions by Greek-speaking Twitter users for a period
of 36 months.  We detected over a million events where seemingly
unrelated accounts tweeted nearly identical content at nearly the same
time.  We filtered these concurrent content injection events and
detected a set of 1,850 accounts that repeatedly exhibit this pattern
of behavior, suggesting that they are fully or in part controlled and
orchestrated by the same software.  We found botnets that appear for
brief intervals and disappear, as well as botnets that evolve and
grow, spanning the duration of our dataset.  We analyze statistical
differences between bot accounts and human users, as well as botnet
interaction with user communities and Twitter trending topics.
\end{abstract}

\section{Introduction}

Twitter's open nature towards user communities, along with its
capacity to disseminate information, makes it very appealing for
malicious entities.
These entities aim to influence public opinion or at least convince
people of their ability to do so, for reasons such as personal
popularity, unsolicited advertising and gaining political influence.
%
The activity of these agents is often organized in the form of
\emph{botnets}, which are groups of \emph{sybil} accounts that
collectively seek to influence ordinary users. 
The percentage of bots among Twitter users, has been estimated to be
between 9\% and 15\%~\cite{varol2017online}.

Although not all automated accounts are malicious, the potential
magnitude of damage has driven a lot of research into the detection of
bot
accounts~\cite{subrahmanian2016darpa,ferrara2016rise,varol2017online},
focusing mostly on English-speaking and Arabic-speaking parts of the
network%
\footnote{Often by machine-translating the latter to the former.}.
A large part of the related literature focuses on detection of bot
accounts, based on attributes such as content, patterns of activity
and relations to other users.
The literature covers a multitude of methods utilized towards this
goal such as simple heuristics, statistical methods, machine-learning
and artificial intelligence.  
However, less work has been published on the systemic grouping of bot
accounts into botnets and the evolution of these botnets over a long
period of time.

This article aims to analyze the behavior of Twitter botnets
over time, within the full Twitter graph.  First, we use the technique
of Concurrent Content Injection Detection
(CCID)~\cite{cao2014uncovering} to identify a set of Twitter accounts
consistently posting \emph{almost similar} tweets, at an \emph{almost
synchronous} time and this occurs \emph{often} over time.
We study the parameters of CCID, i.e., what constitutes \emph{almost
similar tweet}, \emph{almost synchronous time} and how \emph{often}
is unlikely to be a coincidence.  Sections~\ref{sec:jaccardthreshold},
\ref{sec:timethreshold} and ~\ref{sec:coincidencethreshold} present a
study of how these parameters affect the number of detected bots.
Overall, this analysis will select a set of users that have a high
probability to be bot accounts.

Second, we analyze the interactions of these bot accounts on Twitter
to detect how they form botnets that cooperate towards the same
purpose.
We grouped bots into botnets by clustering them on the ``copy events''
graph, where botnets emerge as separate, dense clusters.  Then, we
explore the interactions of these botnets with the rest of Twitter:
\begin{itemize}
\item In the ``tweet'' graph: Do they mention, retweet, quote, or reply to specific communities in the corresponding Twitter graphs?
\item In the ``favorite'' graph: Do they like each other's tweets?  Do they like tweets of specific users or communities?
\item In the ``follow'' graph: Do they follow each other? Do they follow other communities?
\item In the ``list-similarity'' graph: Do many \emph{other} people think they are similar to each other?
\item Feature-based clustering: Do they have similar content-based features? How do they differ from non-bot users?
\item By their interactions with ``trending topics'': How do they interact with trending topics?
\end{itemize}

To perform these tests, we crawl and analyze a dense subset of the
Twitter graph, namely all tweets of all Greek-speaking users of
Twitter, for a period of 36 months.  For each layer of the graph,
i.e., tweets, retweets, mentions and replies, quotes, followers,
favorites, and list memberships, we detect communities of users and
classify them using a large list of exemplar accounts.  Next, we study
the interactions of botnets with user communities in each layer and
show how botnets change together with the rest of Twitter, appearing,
disappearing, and evolving, over a period of 3 years.

In summary, this paper makes the following contributions:
\begin{itemize}
\item We apply an existing method for detecting automated Twitter
activity on a dense, language-based community, demonstrate how to
configure and evaluate its performance, and evaluate our results
against existing bot detection techniques.
\item We discover how Twitter bots interact with users and among each
other using multiple layers of the Twitter graph.
\item We discover how the bots form bot networks and where these are
located inside the Twitter graph.
\item We use multi-layer graph analysis, including layers previously
unused in bot-related literature, such as list memberships and
quote-retweets, to classify bot networks according to their apparent
objectives and areas of interest.
\item We measure how the bot networks evolved over time, as the Twitter graph also evolves.
\end{itemize}

\section{Methodology: Bot Detection}

We used CCID to detect accounts that behave in a way that suggests
automation~\cite{cao2014uncovering}.  CCID is a method to detect
automated accounts based on the assumption that when two or more
accounts perform similar activities in a seemingly synchronized way,
beyond the probability of coincidence, this is a strong indication
that these may be automated and operated by the same piece of
software, or managed by the same person using botnet management
software.

Our system also uses this assumption to discover bots; namely, we
consider a Twitter user to be a \textbf{\emph{bot}} based on the
following criteria: 
\begin{itemize}
\item Do they post \emph{almost similar} tweets with other users?
\item Do they post these tweets at \emph{near synchronous} times with these other users?
\item Do they do it \emph{often enough}, so as not to be a coincidence?
\end{itemize}

We group \emph{almost similar} tweets based on the Jaccard similarity
index between their sets of words.
In order to avoid false positives and minimize false negatives, we
fine-tune the threshold of this similarity index, by performing a
sensitivity analysis described in Section~\ref{sec:jaccardthreshold},
resulting in a selected threshold of 70\%.

We also consider tweets to be \emph{near synchronous} if they fall
within the same sliding \emph{time window}.  
We fine-tune the length of the sliding window by performing a
sensitivity analysis described in Section~\ref{sec:timethreshold};
based on this analysis, we select a time window of 10 minutes, sliding
over 5-minute intervals.

If two or more tweets fall within the same time window and are 70\% or
more Jaccard-similar, we consider them to be part of the same
\emph{copy event}.  
As \emph{copy events} may occur by accident in Twitter, especially
given the small size of tweets, we further filter the copy events
detected, to rule out coincidental events.

We consider copy events between two Twitter users to be \emph{not a
coincidence}, if there are multiple copy events within multiple
intervals and for multiple tweets, between these two users.
We fine-tune the coincidence threshold by performing a sensitivity
analysis, described in Section~\ref{sec:coincidencethreshold} and we
select a threshold of \emph{coincidence}.

The final, filtered result of these criteria is a directed weighted
graph where nodes represent users and edge weights measure the number
of \emph{copy events} between two users.
For each edge, the source node is the original poster and the target
node is a copying user.

For this study, we used a corpus of 720 million tweets crawled between January 2016 and December 2018. 
This corpus includes publicly available tweets from a large percentage
of all Greek-speaking users, i.e., users that have tweeted more than
5\% of their tweets in Greek~\cite{twawler}.
Our crawler fetches tweets, list memberships, follow relations, and
favorites from Twitter's API and stores this information in a local
MongoDB database.
The analyses described below are implemented in python, using the
iGraph library for graph analysis.





\subsection{Jaccard Similarity Threshold Study}
\label{sec:jaccardthreshold}

Even though automatically generated content tends to be identical,
even automatically generated tweets may differ slightly, e.g., due to
two different shortened URLs, or the addition of a prefix or postfix.
To be able to detect similar but not identical content, we use Jaccard
similarity to compare tweets.  Specifically, we compute the Jaccard
similarity between the bag-of-words representation of each tweet and
consider tweets with an above-threshold Jaccard similarity to be
possible copies.  We analyze the sensitivity of this method to the
selected threshold, using a dataset outside the analyzed period,
corresponding to February 2019.

\begin{figure}[t]
\centering
\includegraphics[width=\linewidth]{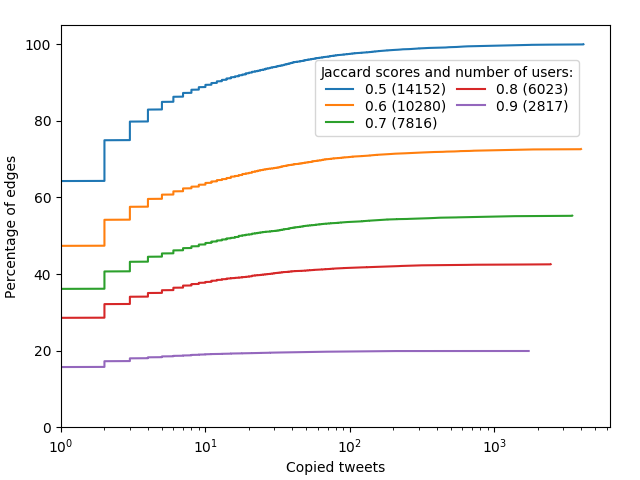}
\caption{CDF of the numbers of copied tweets per user in the sample, for various
Jaccard thresholds.}
\label{fig:jaccardtweets}
\end{figure}

Figure~\ref{fig:jaccardtweets} shows the percentage of users (Y-axis)
that have copied at least a number of tweets (X-axis), for different
thresholds of the Jaccard similarity index. 
For example, at a threshold of 0.5, 14,152 users are involved in
copy events: more than 60\% of them have copied only 1 tweet, whereas
the most active copier, at the rightmost point of the curve, has
copied more than 4,000 tweets.
Raising the threshold to 0.6 reduces noise, although still 50\% of the
users have only 1 copied tweet.

At each incremental step of the threshold up to 0.8, the CDF curve is
following the same shape as the 0.5 threshold curve, whereas noise
(users with 1 copied tweet) is decreased.  Finally, at threshold
levels of 0.8 and 0.9, although noise is significantly decreased, we
also notice that users with the most copy events have lost thousands
of copy events.

\begin{figure}[t]
\centering
\includegraphics[width=\linewidth]{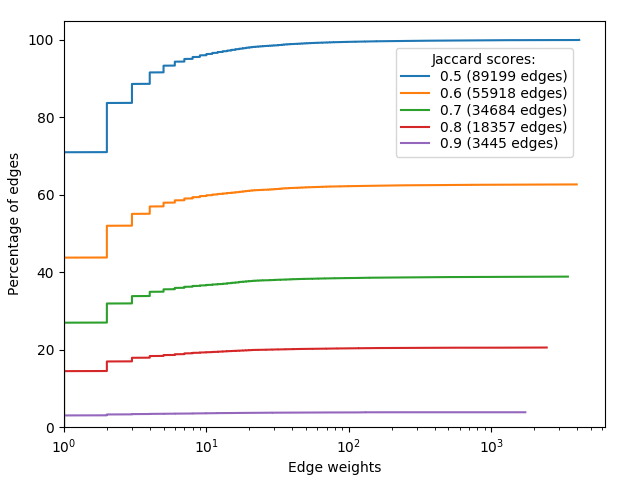}
\caption{CDF of the numbers of copy events per edge, for various
Jaccard thresholds.}
\label{fig:jaccardedge}
\end{figure}

In addition to the distribution of the number of copy events, we study
the Jaccard similarity threshold with respect to the frequency of copy
events detected between pairs of users.  Intuitively, if two users are
frequently involved in separate copy events, we consider that there is
a higher probability they belong to the same botnet.  Conversely, if
two users are only observed to be involved in a single copy event
together over a large period of time, we consider that there is a
higher probability the copy event was coincidental.  We measure this
by forming the copy event graph, where undirected edges among users
are weighted by the total number of copy events involving these users.

Figure~\ref{fig:jaccardedge} shows the distribution of edge weights
in the copy event graph, for different Jaccard thresholds.
At the 50\% threshold, almost 70\% of edges have weight 1, by increasing thresholding we reduce that type of noise. Differences between 70\%, 80\% and 90\% are almost equal, only a step from 60\% to 70\% removes a high portion of the noise where edges with weight 1 are reduced to around 45\% and allows us to keep a high volume of users. It is important to mention that 100\% of user volume is computed by the base of the lowest threshold that we use in our experiment. For that reason every time the threshold is increased, the volume of users is decreasing. This is because a higher portion of tweet text should be the same between two posts.
A summary of those plots offers two options of thresholding: 60\% and 80\%, where both of them have pros and cons. For our implementation, we can't use both of them so we choose the middle solution of 70\% that reduces a high portion of noise and at the same time keeps the high volume of users. 


\subsection{Time Window Study}
\label{sec:timethreshold}
Time can be used as a parameter to determine who is the origin of information and who is the copier.
Twitter API provides event details of each tweet with a timestamp of the event.
Computing Jaccard between all available tweets 
is time-consuming and unnecessary. In case of breaking news, bot accounts are interested in posting that information, as fast as possible, in order to be in a trending topic. According to that, we have performed measurements that identify the best time window within the higher portion of copy events that are happening to have in mind the computational time. At the long time window, computational time can spend an unacceptable amount of time, since we compare all to all tweets within the time window.
Specifically, we tried time windows of size 2.5, 5, 10, 15 and 20 minutes,
over the same one-month data, used in \ref{sec:jaccardthreshold},
which detected around 800K, 1.1M, 1.6M, 1.9M and 2.2M pairs of similar
tweets, in total over all the time windows of the dataset,
respectively. Computation time was around 2, 4, 9, 18 and 34 hours,
respectively, on a single 8 core machine, for the window durations
listed above. Assuming the results, we notice that the 10-minutes window identifies a significant portion of the events with lower computational time step with a comparison of the 15-minutes window.  Based on this trade-off, we selected a time window of 10
minutes, sliding over 5-minute intervals for the complete dataset.

In our evaluation scheme, between these time windows, 10 and 15 have a difference of 0.3M pairs.
One can assume that by choosing a 10-minute window our implementation will lose those events.
That assumption is not accurate, because we utilize a sliding time window of \emph{N/2}, where \emph{N} is the size of the time window ~\cite{cao2014uncovering}. It means that we keep a connection between events in different time windows. Within a 10 minute window with
a sliding window of 5 minutes, we can equally capture the copy events that happen in 15 minutes and it is closer to the idea of almost concurrent than any other larger window.

\subsection{Detecting concurrent content injection}
\label{sec:ccid}

In order to detect synchronized activity, we use a standard rolling window
method to segment the very large body of
tweets into smaller chunks so they can be analyzed in parallel and
independently, consisting of all tweets in a time window. We overlap
the $N$-minute windows by $N/2$-minutes, so that each tweet is
visited twice, independently. Each time window's worth of tweets
(task) is analyzed by mapping each tweet into a bag-of-words, using
the standard NLTK tweet tokenizer that simplifies punctuation and  uses each URL and hashtag as a single word. We found URLs specifically to be important, as the
same content is often injected by changing simply the shortened
URL in the tweet into another.
For each 10-minute window in the dataset, we compare the bag-of-words
of each tweet with every other tweet's bag-of-words in the same window,
using the standard Jaccard similarity:
$$ [T_i \cap T_j] / [T_i \cup T_j] = S_{ij} $$
where $T_i$ is the set of words in tweet $i$ and $S_{ij}$ is the
the similarity between the two tweets. 
\begin{table}[!t]
\centering
\begin{tabular}{|c|c|c|c|}
\hline
Year &  Involved Users  & Source Only & Copiers \\ \hline
2016  & 44,312 & 9,645 & 34,667 \\ \hline
2017 & 30,628 & 7,698 & 22,930 \\ \hline
2018 & 24,670 & 6,279 & 18,391 \\ \hline

\end{tabular}
\caption{Statistics of users involved in copying event graphs. That table presents the number of users that are represented in our graph as source only (those accounts produce original tweets and they are never copying tweets) and also here presented copiers users (that users are copying tweets in time of year).}
\label{tab:involved}
\end{table}
If the Jaccard similarity of the word sets of the two tweets exceeds
a threshold of 70\%, we consider them to be similar and mark them
accordingly. Then, each group of similar tweets in the task, form a
single ``concurrent content-copying'' event, where the first tweet is
considered to be the source and all subsequent tweets to be copied.
Finally, the result of each task is a set of such events, comprising
of the user account IDs for the author of the source tweet and the IDs
for the users that copied it within the 10-minute window. We index
the copying events by source tweet ID, to avoid double counting of
events that occur within the same 5-minute window, analyzed within two
overlapping 10-minute windows.
The resulting copying events are also stored in the crawler's MongoDB,
enabling the further analysis to easily query them together with the
crawled follow and list similarity graphs, residing in the same
database.
To schedule all detection tasks, each task analyzing the tweets of a
10-minute window, we first index the tweet corpus by timestamp and
use a simple FIFO scheduling algorithm to map 10-minute windows of
consecutive tweets to tasks, running on a small cluster of 5 nodes
with 8 cores and 20GB of memory per node. Since even a 10-minute
window can have a large number of tweets in peak times and the
comparison computation is quadratic, we do not observe I/O to be a
bottleneck for a single MongoDB store. The analysis took 15 hours to
complete, with all nodes being used in full CPU capacity during all
that time, except the last hour.

The analysis detected 3,704,910 coincidence events within 
all 10-minute windows for three years (2016-2018 Table ~\ref{tab:involved}).
In total, there are 71,669 users involved. These include 55,735 
''source'' users that were the first to tweet
something that was copied within a 10-minute window
and 56,019 ''copying'' or ''reposting'' users.
Of these, 15,644 users are source users but never copying users.

\subsection{Number of Events Study}
\label{sec:coincidencethreshold}

\begin{figure}[h]
\centering

\begin{subfigure}{.5\textwidth}
  \centering
  \includegraphics[width=\linewidth]{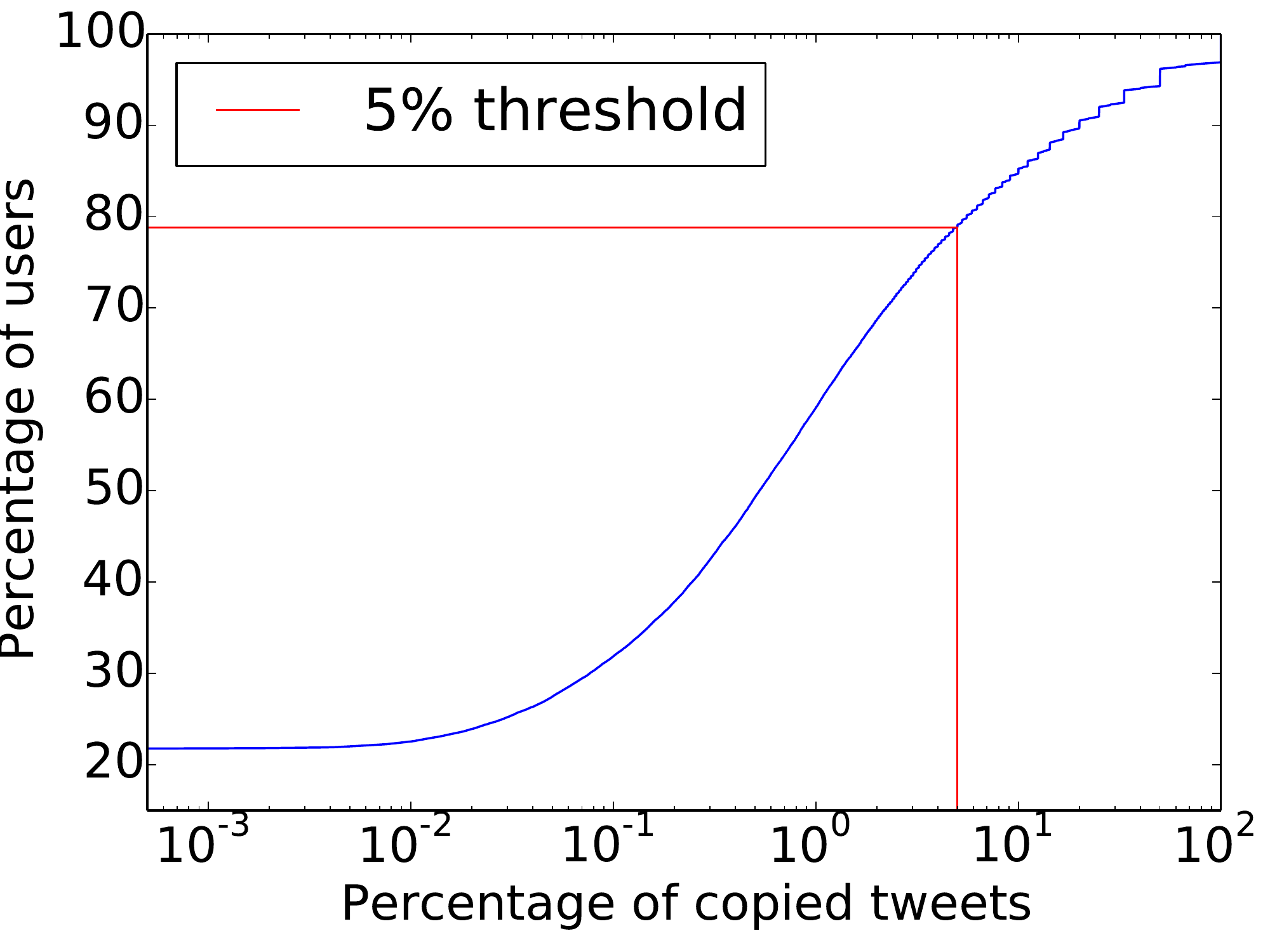}
  \caption{Copied tweets filtering distribution.}
  \label{fig:sub1}
\end{subfigure}%
\begin{subfigure}{.5\textwidth}
  \centering
  \includegraphics[width=\linewidth]{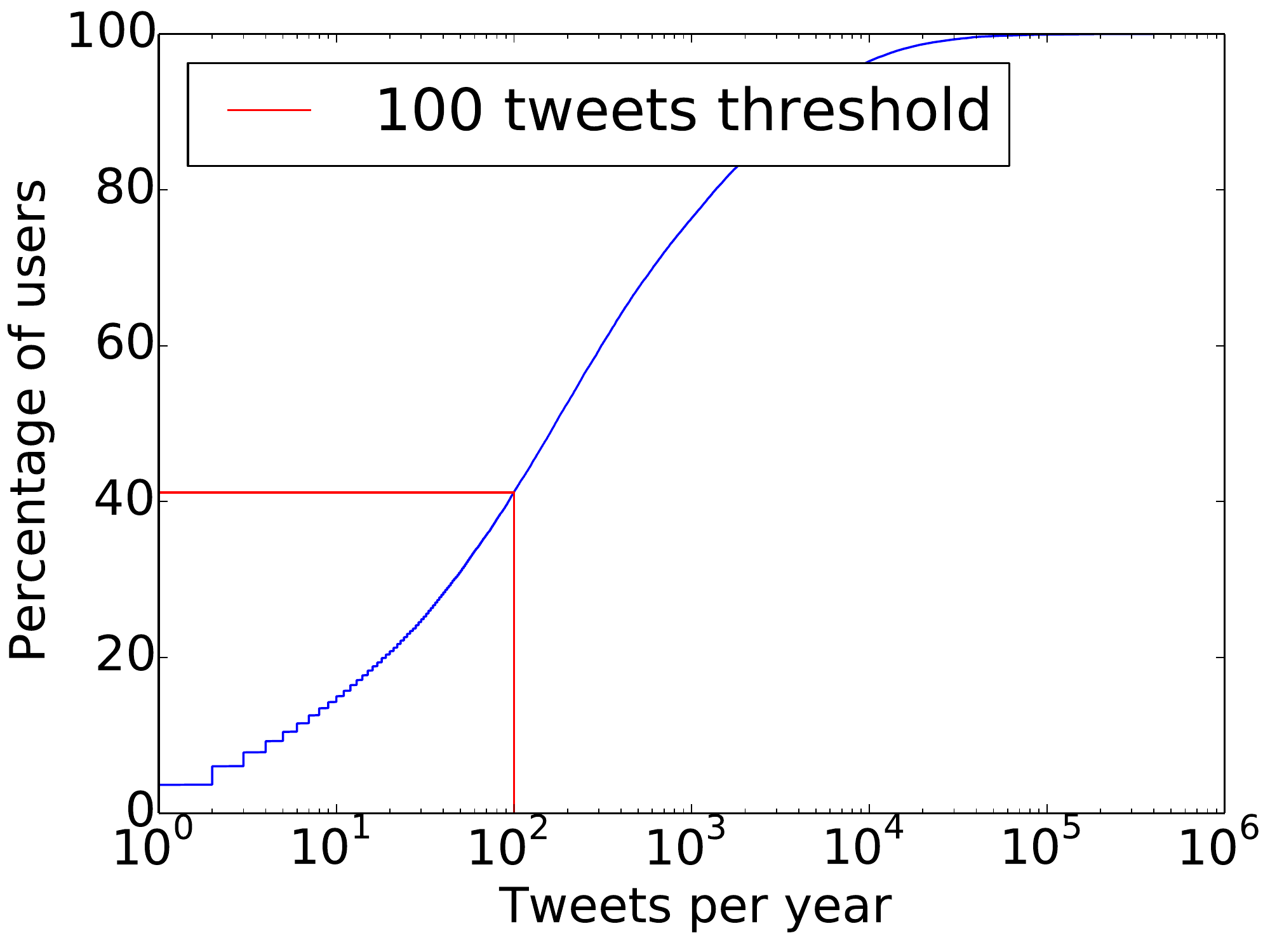}
  \caption{Number of year tweet distribution.}
  \label{fig:sub2}
\end{subfigure}
\caption{This figures show the distribution of users (a) by number of copied tweets and (b) by number of tweets per year. Those parameters are used in order to filter the graph and reduce the false positives.}
\label{fig:test}

\end{figure}

Via manual analysis, we identify a portion of users that copied an insignificant percentage of tweets that can be considered a coincidence.
Also, we identify some of the accounts with minor activities volume, that does not fit at \emph {bot activity pattern}.  In order to remove such \emph {noise} we manage to filter our graph by a fine-tuning parameter of the copied tweets percentage. To explore the right cutoff threshold, we extract a graph $G_{icc}^T$ for each of several threshold levels. Table~\ref{tab:percent}  shows for several cut-off thresholds, the characteristics of the filtered graphs where the first column shows the copied events threshold percentage. The second
column shows the number of accounts that have more copied tweets than the given threshold. The third column shows the total number of edges that remains after the filtering. Finally, the last column represents the number of connected components. For both of the parameters we generate the distribution. In figure~\ref{fig:test} we present the amount of users that are removed by each filtering method.\\

The resulting graph consists only of users that have posted at least
100 tweets, of which at least 5\% are copied tweets.  We compute such
a graph for each calendar year in the dataset, as well as the total
graph for the 3 years.  

\begin{table}[!t]
\centering
\begin{tabular}{|c|c|c|c|}
\hline
Threshold $T$\% & \#Accounts & \#Edges & Connected Comp.\\
\hline
\hline
0\% & 38,541 & 189,195 & 1860 \\
\hline
1\% & 21,482& 111,428 & 398 \\
\hline
3\% & 11,800 & 66,037 & 210 \\
\hline
5\% & 8,496 & 49,735 & 156 \\
\hline
10\% & 5,549 & 32244 & 107 \\
\hline
\end{tabular}
\caption{
Size of filtered graph for different threshold levels.
Experiments performed on 2016 graph data. Number of nodes with 0\%
threshold are differ with Table:~\ref{tab:involved}, because we
applied already the filtering step of removing not active accounts. With that step we already
removed users with less than 100 tweets per year.}
\label{tab:percent}
\end{table}

\subsection{Bot graph}
From the set of copying events detected, we extract the
\emph{concurrent content injection graph} $G_{cci}^5$, where vertices
are user IDs and weighted directed edges signify concurrent content
injection by the two users.
To construct the graph, we scan all copy events; for each copy event
we create a set of directed edges from the source (first time when the tweet was seen at time window) to each copying
user (following occasions of similar tweet), but not between the copying users.  That procedure generates multi-edges between the same pair of users in the case when the same account/bot often post/copy tweets from the same origin. We reduce multi-edges into weighted edges graph form, where edge weight represents the number of copy events.

The steps we follow to filter the graph are:
\begin{enumerate}
\item For each user, calculate the percentage of user's tweets we found to be copied
\item Disregard the users for whom copies are lower than a threshold T of their total posts
\item Remove not active users with less than 100 tweets per year and create the final graph per year of data
\item Perform community detection and strongly connected component analysis for each $G_{cci}^T$
\end{enumerate}
We develop a 3-years graph $G_{cci}^5$ that contains 15,433 nodes involved in copy events and 239,916 directed edges. Table~\ref{tab:allgraphs} presents the concurrent content injection graphs 2016, 2017, and 2018 separately. 

\begin{table}[!t]
\centering
\begin{tabular}{c@{}c@{}c@{}}
& Origin & After filtering \\
\begin{tabular}{|r|}
\hline
Year \\ \hline
2016 \\ \hline
2017 \\ \hline
2018 \\ \hline
3-years \\ \hline
\end{tabular}
&
\begin{tabular}{|r|r|}
\hline
Nodes & Edges\\ \hline
44312 & 240600 \\ \hline
30628 & 148276 \\ \hline
24670 & 121637 \\ \hline
  &  \\ \hline
\end{tabular}
&
\begin{tabular}{|r|r|}
\hline
Nodes & Edges\\ \hline
8496 & 49735\\ \hline
7404 & 46580\\ \hline
5601 & 32911 \\ \hline
15433 & 239916 \\ \hline
\end{tabular}
\end{tabular}
\caption{Number of nodes/edges
in graph $G_{cci}^5$ before and after filtering, for each year and for
the complete dataset.}
\label{tab:allgraphs}
\end{table}

\section{Analysis of Bot Activity}

\begin{figure}[t]
\centering
\includegraphics[width=\textwidth]{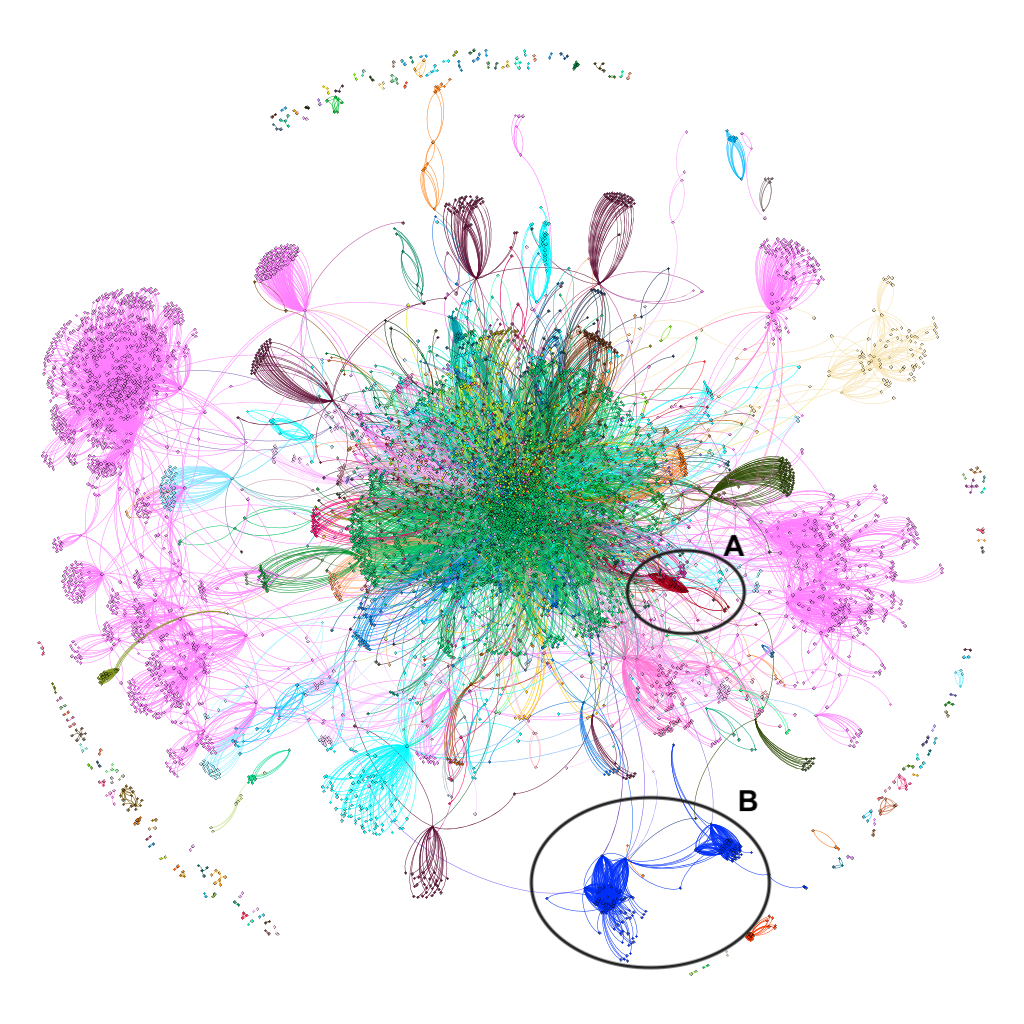}
\caption{Bot graph of 2016}
\label{fig:copied16}
\end{figure}

\begin{figure}[t]
  \centering
\includegraphics[width=\textwidth]{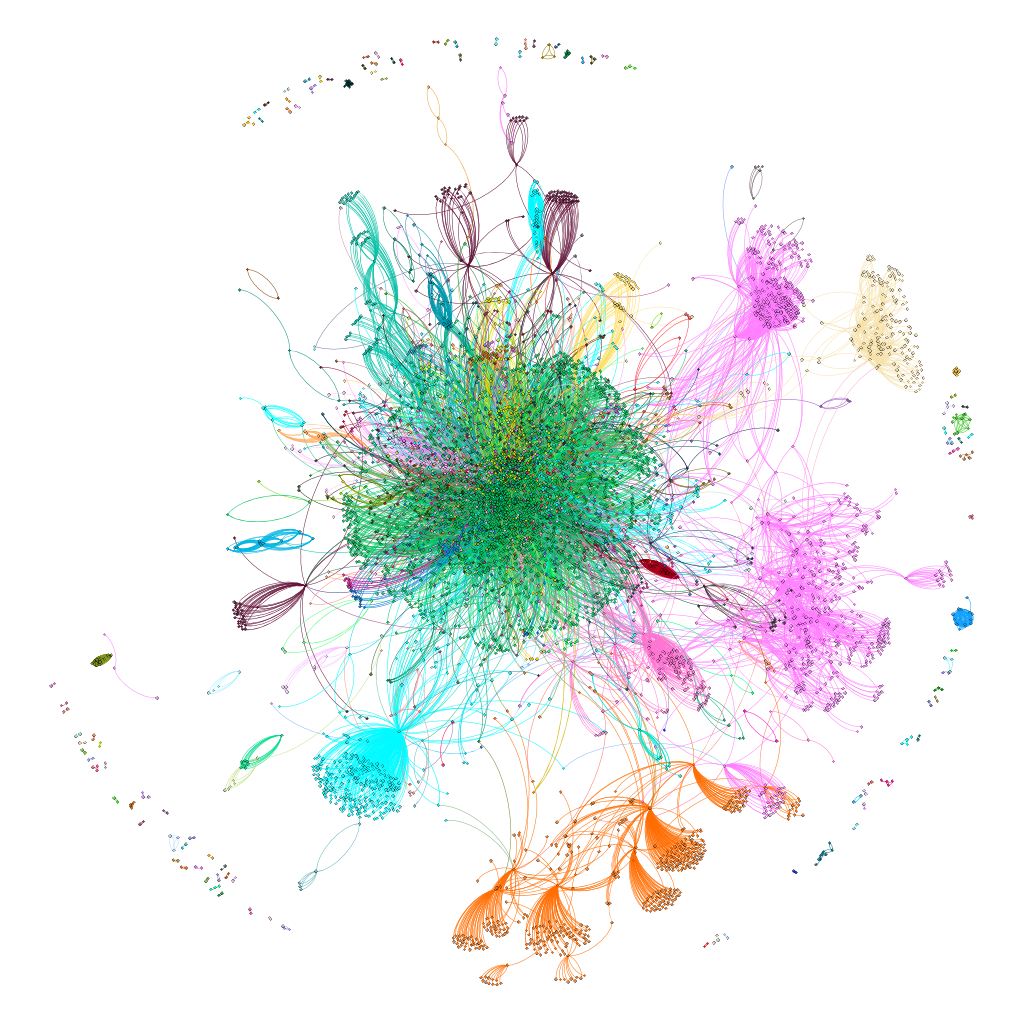}
\caption{Bot graph of 2017}
\label{fig:copied17}
\end{figure}

\begin{figure}[!b]
  \centering
\includegraphics[width=\textwidth]{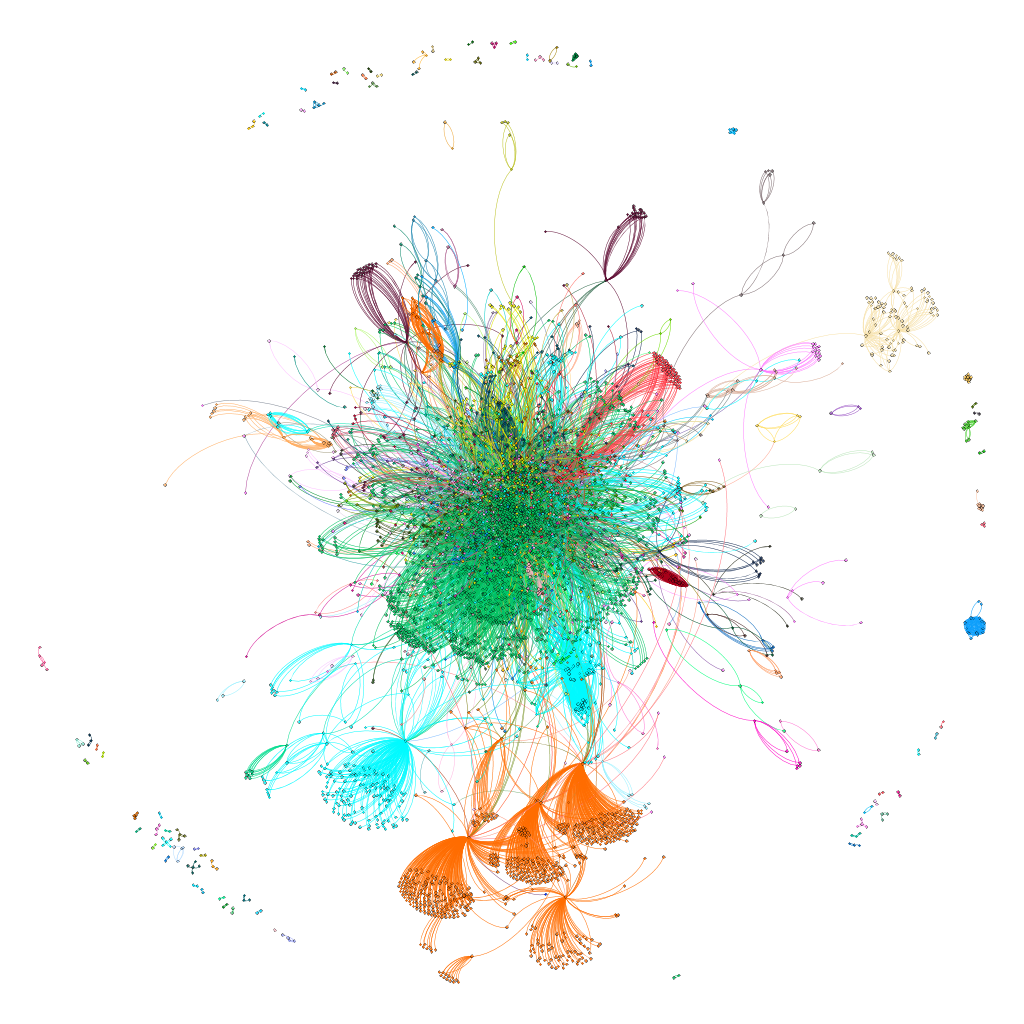}
\caption{Bot graph of 2018}
\label{fig:copied18}
\end{figure}
The developed procedure, identifies post/tweet events with similar content between
users within the Twitter social network. By filtering those events, we
reduce the number of users that may be considered by accident as bots.
As a result of the implemented procedure, we establish multiple graphs that represent the activity of each year independently (figures: \ref{fig:copied16}, \ref{fig:copied17} and \ref{fig:copied18}).
Those graphs visualize the relation of who copies/steals from whom. Additionally, an interesting research field is to develop a methodology that identifies the group of accounts that are collaborating in groups/cliques, also known as a botnet. Accounts that are involved in such groups/botnets are performing similar/coordinated actions. Our methodology not
only identifies such botnets via \emph {copied event graph} but also 
manages to observe the evolution of those botnets over a 3 years
period.
Figures \ref{fig:copied16}, \ref{fig:copied17}, and \ref{fig:copied18}
show the $G_{cci}^5$ for 2016, 2017, and 2018, respectively.

\subsection{Botnet detection}
To identify coordinated activity between the detected bots, we perform
community detection on the $G_{cci}^5$ graphs. Our objective is to
study the evolution of botnets over time. Hence, for the detected
botnets to be consistent and relatable among the yearly graphs, we
perform community detection for the whole 3 year period. To do that,
we merge the three-yearly graphs into one total $G_{cci\_total}^5$ graph which contains all detected bot accounts and the connection between them. Thereafter, we perform community detection on $G_{cci\_total}^5$ and then project the detected communities back onto the yearly datasets.

Figures \ref{fig:copied16}, \ref{fig:copied17}, and \ref{fig:copied18}
are presenting the resulting communities with consistent colour coding and location of bot accounts. In order to generate those figures, we used the modularity algorithm in Gephi~\cite{gephi} over the total $G_{cci\_total}^5$ graph. For a rendering procedure of the full graph, we are using the \textit{Fruchterman Reingold} graph visualization
algorithm in Gephi. Finally, we projected the total rendering into
yearly slices, where the same user involved in copy events is placed
in the same position. Note that figure \ref{fig:copied16} has two
components/botnets indicated by (A) and (B) circles. The accounts located within each component were created at the same date (figure \ref{fig:botcreationdates}), with a similar pattern of behaviour (explained later in Section: \ref{sec:featurepaterns}). Due to the fact
that accounts/bots will be placed in the same position through the
different years, we can compare figure \ref{fig:copied16} and
\ref{fig:copied17}, where the botnet indicated by (B) disappears and
the botnet indicated by (A) remains active.

\subsection{Feature patterns}
\label{sec:featurepaterns}
\begin{figure}[h]
\centering
\includegraphics[width=\linewidth]{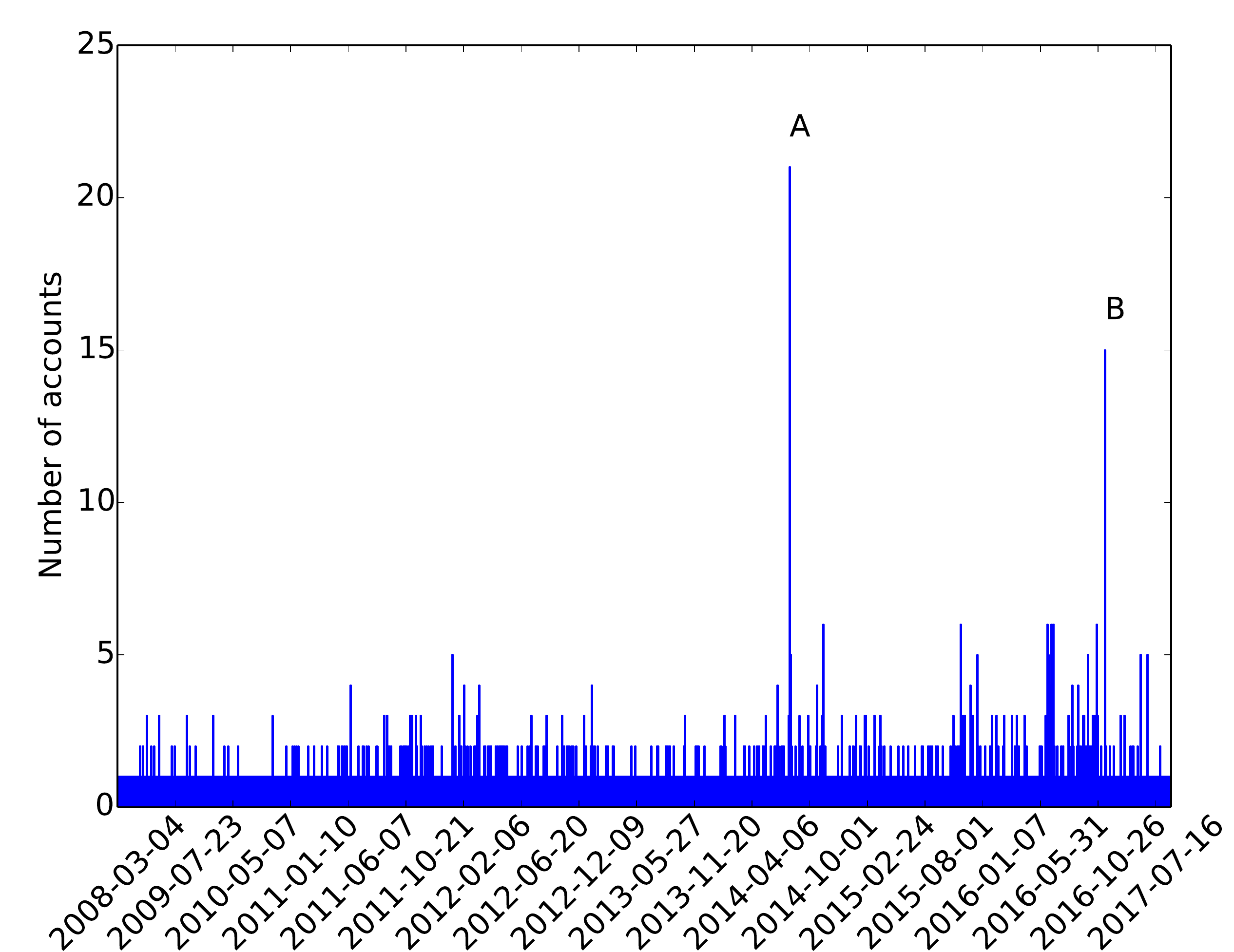}
\caption{Distribution of dates of creation for accounts in $G_{icc}^{15}$}
\label{fig:botcreationdates}
\end{figure}

After filtering the graph as described above, we extracted a set of
hundreds of features~\cite{twawler} for the 1.850 bots and 20.000 normal accounts. With the use of Lasso feature selection, we managed to identify the features for those two categories of accounts. In the appendix section \ref{appendix}, we are presenting some of them. It is important to mention that such features indicate just the differences in the distribution between normal and bot accounts and not necessarily all of them can be used in order to classify an account as a bot or not.\\
Through manually analysis we identify interesting patterns within the features; for example, figure ~\ref{fig:botcreationdates} shows the distribution
of the dates these accounts were created. The two spikes highly
correlate with two small clusters of almost equal size that are
visible in figure ~\ref{fig:copied16}.
Here we notice that the first group is drawn in
red and indicates the (a) circle
and the second group is drawn in blue colour and indicates the (b)
circle.  Through manual analysis, we identified
that the first group of accounts, created in the same date, uses Greek
university names for their account names.
Apart from the user names, we distinguished a pattern of profile pictures and website templates:  
all of these accounts use the same template for the website that they are referring from.
These accounts are reposting the same announcements, published by the university from which they are using their usernames, in order to lure
students.

The next group of users are the ''bet'' accounts with a simple pattern of
username creation: get a popular football team name and add the word "bets".
Manual analysis of these accounts shows almost identical content in their tweets, even same spelling mistakes in the Twitter body. The content posted by these accounts is directing users to the same website. These accounts seem to be dump bots that are recycling identical content between various accounts without even changing the URL shortener.
Most of these accounts seem to prefer automated services for posting many of their tweets, sometimes synchronously. 
The platforms that were used to post these copy tweets are: now-defunct \texttt{tweeterfeed.com} ({\raise.17ex\hbox{$\scriptstyle\mathtt{\sim}$}}9 million tweets), \texttt{dlvrit.com}, \texttt{tweetdeck}, Facebook's Twitter app, \texttt{ifttt.com}, \texttt{hootsuite.com}, LinkedIn's Twitter app and the ``old tweets''app \texttt{www.ajaymatharu.com}.


More interestingly, Twitter's web client is reported as the source of 384 thousand copied tweets;
these are either false positives or automated tweets trying to resemble users of the web client. 

Moreover, 272 thousand tweets
originating from only 19 users have as source a single URL pointing to the
same website; all 19 involved accounts have female usernames and
avatars of young women reported being stock photographs by Google
image search. These tweets promote the same web page that hosts
mostly content copied from ``legitimate'' news media along with the
administrators' political views. The source application reported for these tweets links to the same website. Curiously, although the
operators of the accounts tried to have them resemble female users,
they did not hide the source application posting these tweets; of
course, this source application is not visible to the interface seen
by most users.

Additionally, almost 20\% of the analyzed accounts have on average a single
URL per tweet, 22\% have at least one in every tweet, while 18\% never
tweet URLs. The median number of URLs per tweet is one, for 60\% of
the analyzed accounts. In combination with the source application
results above, we conjecture that most of the automated content
observed amounts to ``click farming'', \emph{i.e.}, automated scraping
of news stories from legitimate sources, reposting on a blog or fake
news web page and concurrent automated promotion of the resulting
feed-in Twitter, aiming to attract traffic and eventually advertising
revenue.

\subsection{Bot evolution}

Our dataset collection offers an opportunity to observe the evolution of botnets through a 3-years period. We perform community detection where each user is located at the same position and the community colour remains the same during our observation period. In the previous section, we mostly concentrated on two botnets (figure ~\ref{fig:copied16}) with labels A and B. In this section we will generalize the evolution of botnets. In close comparison between the first-year graph, shown in figure ~\ref{fig:copied16} and figure ~\ref{fig:copied17}, where data have the difference of a single year, we already notice differences between the botnets. One of the differences is the \emph{pink} component that is located in a significant portion of the 2016 graph and at the next year this component almost disappears. Another noticeable change is the \emph{orange} botnet that does not exist in the 2016 graph and appears in the 2017 year graph with a larger number of accounts. This particular botnet is growing by adding more accounts at 2018 graph (figure ~\ref{fig:copied18}).\\
A closer look reveals detailed aspects of the bot community evolution, where some botnets that did not exist before, appear and have continuous growing size. Some others are active for a shorter period of time, but our hypothesis is to show that those accounts remain active until they perform some sort of actions (for example spam spread or some type of advertising campaign). The last type of botnets are consistent botnets, where accounts remain active for a longer period of time. Those 3-year graphs show that the botnet evolution is similar to the human communities where some of them are growing, exist for a smaller period of time and communities that remain consistent.

\subsection{Interaction with trending topics}
Twitter curates and maintains a list of trending topics at various levels of locality (international, per region, country, interests,
\emph{etc.}). As trending topics are displayed to all users and often draw user attention, clicks and views, they are often the
target of campaigns aiming at that attention. In that context, we compare the hashtags used in the tweets posted by the 1304 bot
accounts against the trending topics in the 24-hour periods before and after each tweet. We found 500 bot accounts that posted hashtags in
their concurrently injected tweets. Of those, 152 accounts posted
hashtags that occurred in the trending topics within the previous or
next 24 hours. Specifically, 82 accounts posted hashtags that were
found in the trending topics within the previous 24 hours; of those, 49
posted hashtags that were not observed to be trending in the next
24 hours. Conversely, 54 accounts were found to post hashtags that
were observed to be trending in the following 24 hours; 16 of
those posted hashtags that were not trending in the previous 24 hours.

Although not applicable to all bots, this analysis indicates that they
often interact with trending topics, either to hijack user attention
from a legitimately trending topic, or while trying to
astroturf~\cite{ratkiewicz2011detecting} a topic of interest into the
trending topics.

\subsection{Classification using Twitter Interactions}\label{twlist}

Towards a better classification of the topics of interest and communities targeted by
synchronized content injection, we take advantage of the crawled graphs that capture
the interactions taking place between Twitter users. We refer to interactions 
as the actions the users are performing, meaning the follow action, retweet, 
favorite, mention, reply, quote, and placing users in lists. For these actions we gathered 
the graphs that represent these actions, for all the Greek speaking Twitter.

Each action is represented by a graph as described below: 

\begin{itemize}
	\item A follow graph models the relation of who follows whom, into a directed graph.
	\item A retweet graph models the relation of who retweeted whom and how many times, into a directed weighted graph.
	\item A favorite graph models the relation of who favoured whose tweet and how many times, into a directed weighted graph.
	\item A mention graph models the relation of who has mentioned whom and how many times, into a directed weighted graph. 
	\item A reply graph models the relation of who has replied to whom and how many times, into a directed weighted graph.
	\item A quote graph models the relation of who has quoted whom and how many times, into a directed weighted graph.
	\item A list graph models the relation of who is placed with whom in the same lists and the number of lists they are together, into a bi-directed graph.
\end{itemize}

Next, we used all these graphs to classify bot accounts according to the communities they target. 
To discover all the communities we used the igraph ~\cite{igraph} library 
to reveal all the clusters, formed on all these graphs.

Following, we gathered a set of hand-curated exemplar account lists, to characterize each of the
resulting clusters, according to how many exemplar nodes each category contained. 
We used four main exemplar categories: politics, celebrities, news \& media, and brands.
Note that Twitter celebrities may not only correspond to real-life celebrities, 
but are rather Twitter accounts with very high follower counts. We used Wikipedia lists of
brands, journalists, news media, politicians, political parties, \emph{etc.}, to compile 
exemplar lists and mentions in news blogs and media for Twitter celebrities.
In total, there are 639 exemplar accounts, 356 political, 49 celebrities, 
197 news \& media and 37 brand user accounts. Brand and celebrity accounts are less in number, compared
to others because the dataset is restricted only to the Greek community.

\begin{figure}[ht]
\centering
\includegraphics[width=\textwidth]{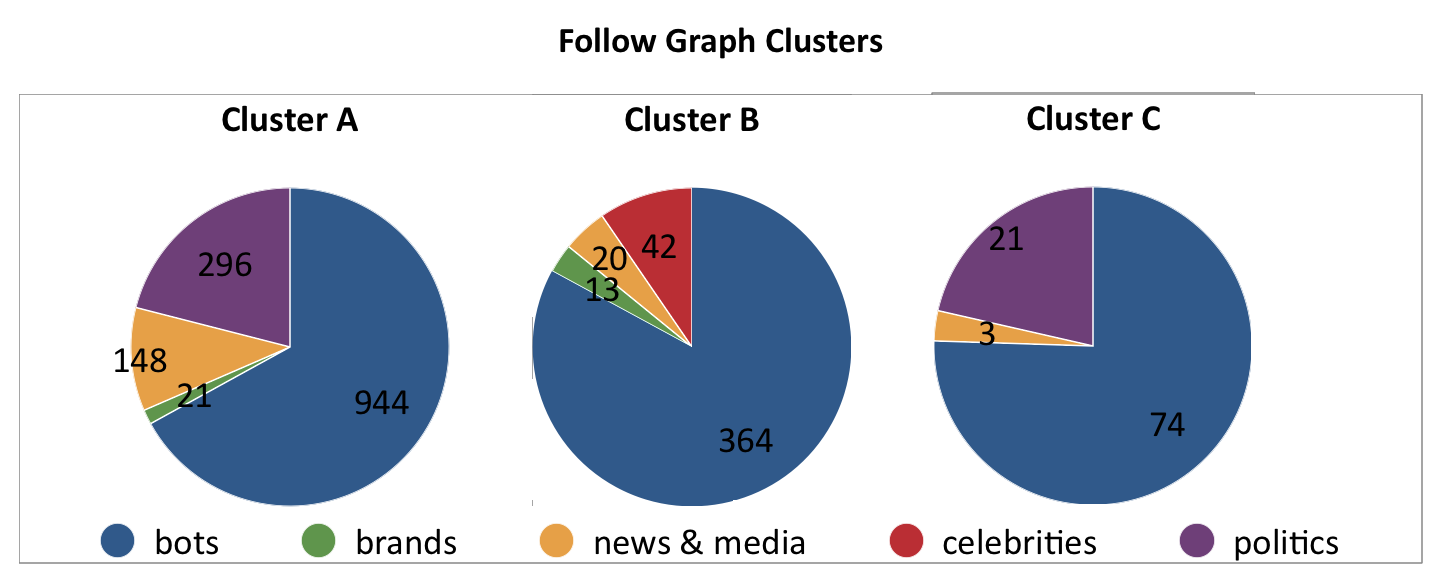}

\includegraphics[width=\textwidth]{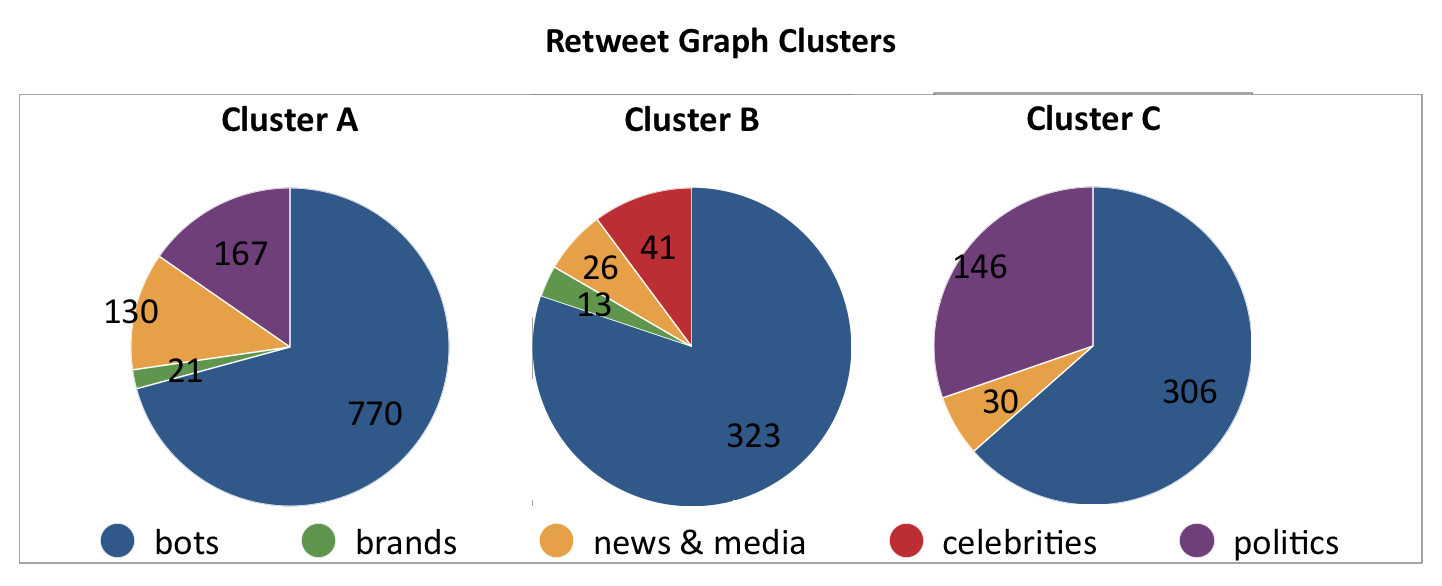}

\includegraphics[width=\textwidth]{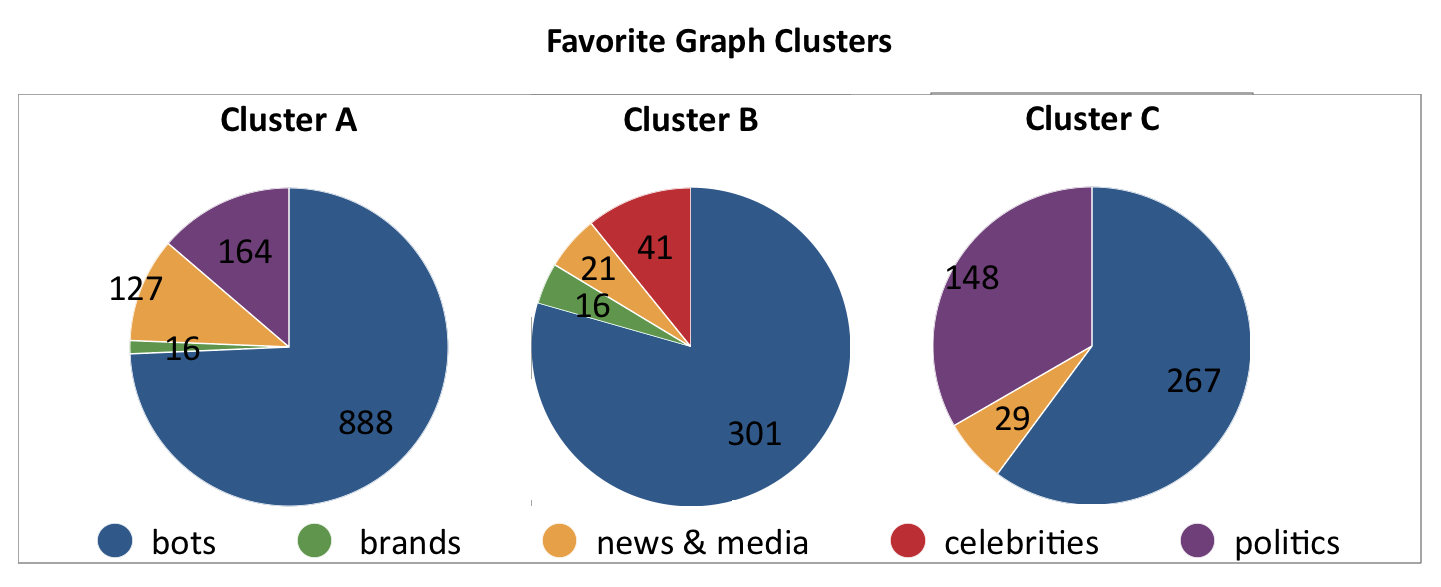}
\caption{Illustration of the 3 top formed clusters for the follow,
retweet, and favorite graphs and the number of exemplars and bots in them.}
\label{fig:communityDetection}
\end{figure}

\begin{figure}[ht]
\centering
\includegraphics[width=0.93\textwidth]{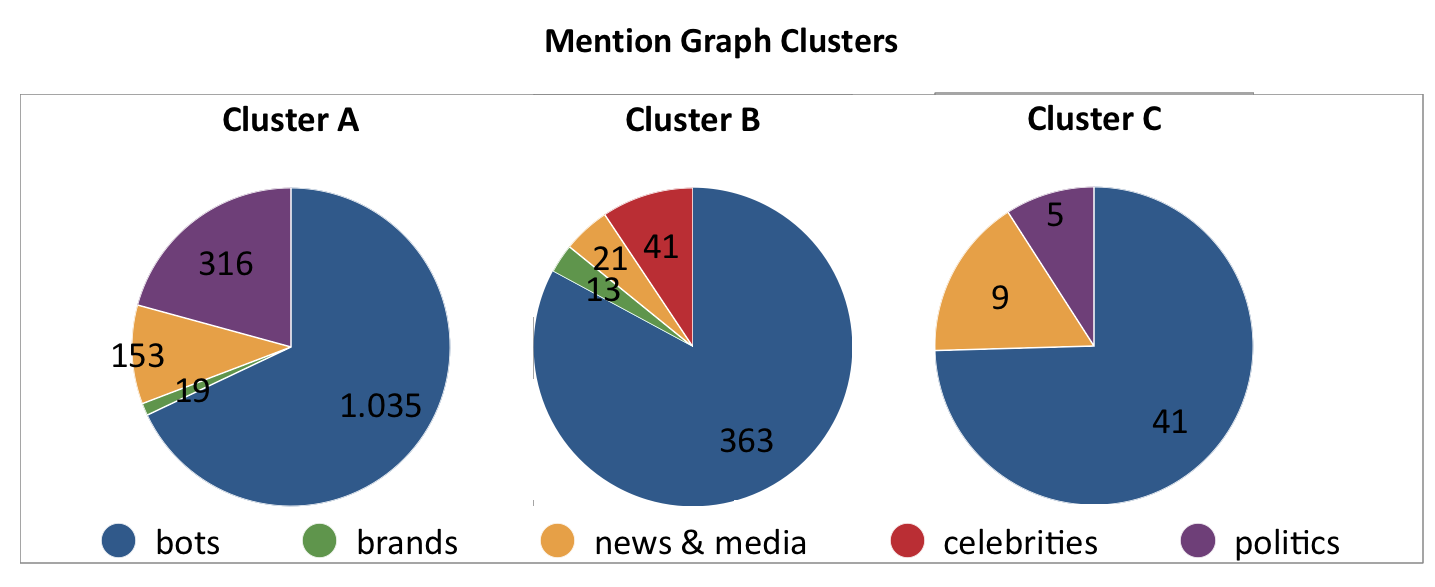}

\includegraphics[width=0.93\textwidth]{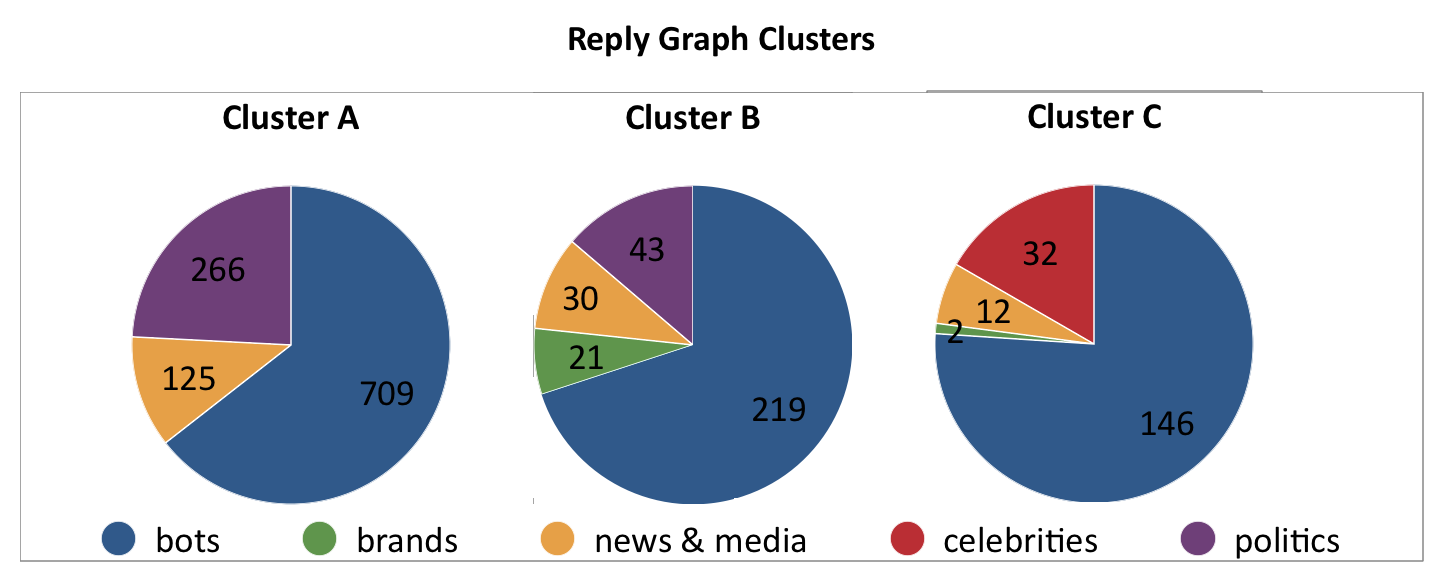}

\includegraphics[width=0.93\textwidth]{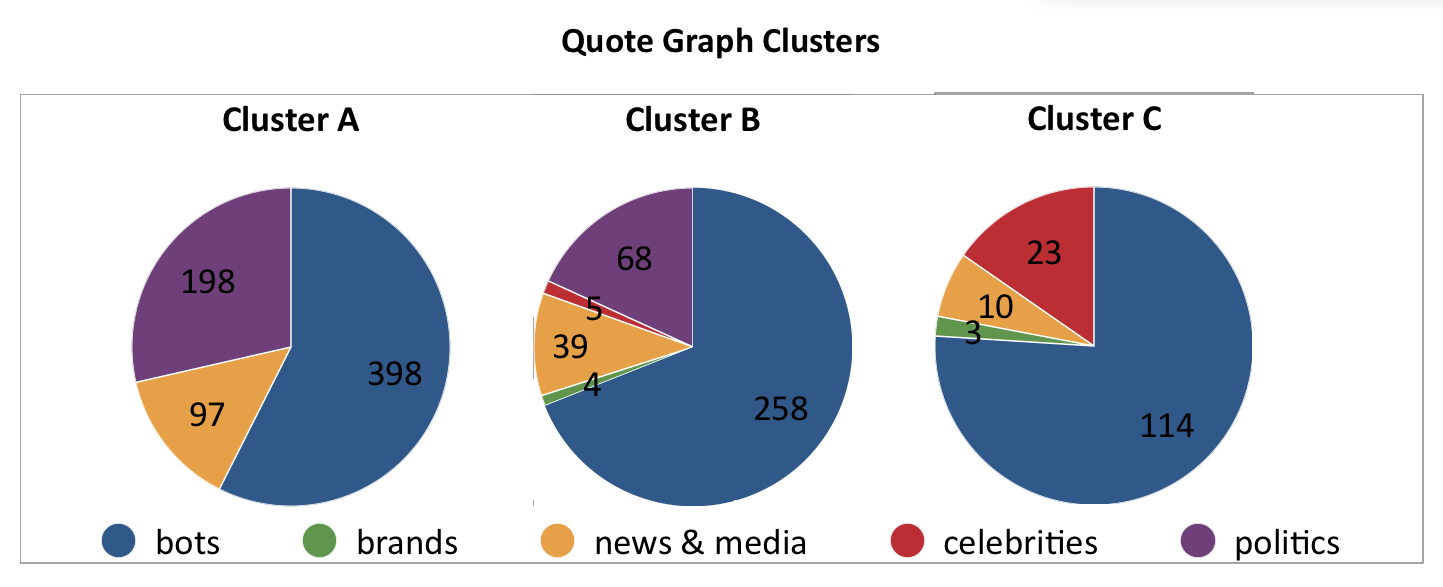}

\includegraphics[width=0.93\textwidth]{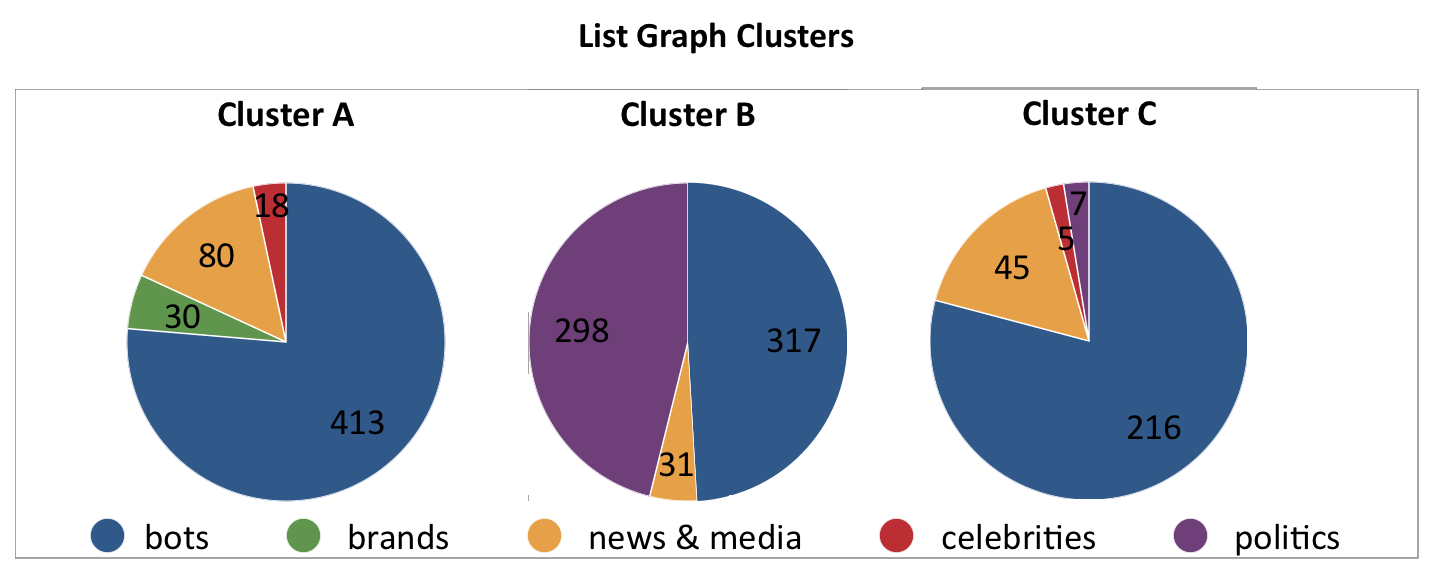}

\caption{Illustration of the 3 top formed clusters 
for the mention, reply, and quote graphs and the number of exemplars and bots within them.}
\label{fig:communityDetection2}

\end{figure}

Figures~\ref{fig:communityDetection}, \ref{fig:communityDetection2}, show for each graph the three most interesting resulting clusters; for each cluster the number of bots that are included, as well as the exemplars of each category.
We consider a cluster interesting when bots and exemplars are included, but due to illustration reasons 
we depicted only three. For example at the top of figure~\ref{fig:communityDetection} we have the 
three most interesting Follow Graph Clusters. Cluster A includes 944 bot users out of 1850, 21 brand users, 
148 news \& media and 296 political users. Cluster B includes 364 bot users out of 1850, 13 brand users, 
20 news \& media and 42 celebrity users. Cluster C includes 74 bot users out of 1850, 3 news \& media 
and 21 political users. We can infer from these three clusters on the Follow Graph that the majority of the bots are mostly 
involved with the political and news \& media accounts and secondary with the brand and celebrity accounts.



From the figures we can assess that most of the bot accounts target mainly 
the clusters that contain the politicians, news \& media users and less 
associated with the brand and celebrity accounts. 

\begin{figure}[ht]
\centering
\includegraphics[width=\textwidth]{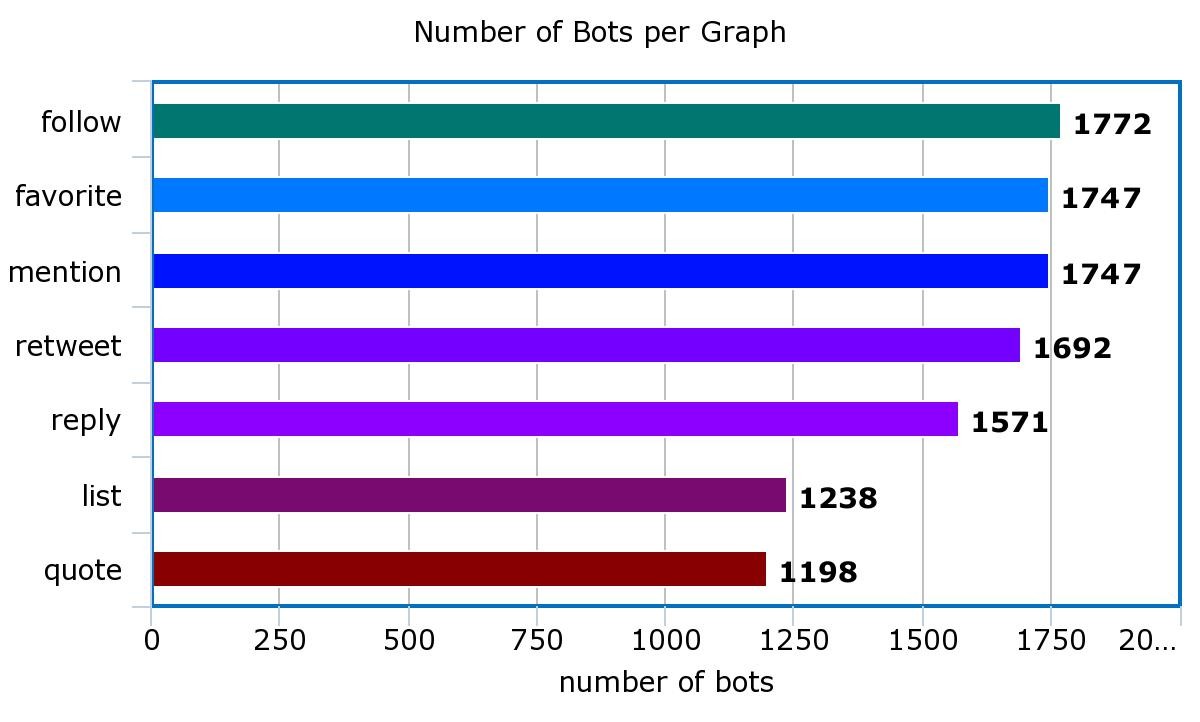}

\caption{Illustration of number of bots found per Twitter action graph.}
\label{fig:totalBotsPerCommunity}
\end{figure}

We can also infer from figure ~\ref{fig:totalBotsPerCommunity} that the engagement of bots in different graphs differs.
In the list and quote action graph we can see that only the 65\% and 67\% of the total amount of bots is engaging and for the rest of the graphs more than the 85\% of the bot accounts is active.

\section{Related Work}

A large corpus of research focuses on detecting and studying botnets
on social media and Twitter specifically.
A botnet is a set of accounts controlled by a single agent. It may run
on single or multiple machines, controlled by a single command and
control infrastructure~\cite{strayer2008botnet}. This structure can
be coordinated, in order to attempt an online attack. Specifically, in
social networks, a botnet can make use of a group of fake accounts in
order to spread malicious content~\cite{thomas2013trafficking}.
Large Twitter botnets have been detected using various techniques,
including the Koobface botnet~\cite{thomas2010koobface} and
others~\cite{thomas2011suspended,scibot,botwalk,cyberbot2017}. Malicious botnets can also be
formed by compromising legitimate user accounts, increasing their
impact~\cite{zangerle2014sorry}.

Detecting bots and botnets is an active area of research, with
state-of-the-art solutions using various
features~\cite{ferrara2016rise}, together with
supervised~\cite{subrahmanian2016darpa} or unsupervised training. BotOrNot~\cite{davis2016botornot,varol2017online} is an online
service that predicts the probability of a Twitter account being a bot,
using a set of more than one thousand features and an array of
classifiers. Using BotOrNot the authors classify the detected bot accounts according
to their content, into spammers, self-promoters and ones that post content from connected
applications. We compare our findings against BotOrNot in
Section~\ref{ctrw}. In ~\cite{chavoshi2016debot} they group Twitter accounts using a novel, lag-sensitive hashtag method based on their warping
correlations. The DeBot unsupervised method detects thousands of bots daily, in near real-time.

Freitas \emph{et al.}~\cite{freitas2015reverse} explore how a network
of social bots can infiltrate the Twitter social graph, by creating
120 bots that span the bot behaviour design space. They show that Twitter
did not filter automated behaviour and that their social
bots were able to infiltrate the real social graph. Zhang \emph{et al.}~\cite{zhang2011detecting} analyze 106,573 Twitter
public accounts by detecting the automation in tweeting. They study
the timing graphs and notice that timing automation fits better to
non-human users on Twitter. In comparison, we extend timing analysis
to much larger datasets and propose ways to filter and understand the
resulting graph, producing better insights as to the goal of the
detected botnets. Edwards \emph{et al.}~\cite{edwards2014bot} use surveys to measure
interaction metrics of users between bots and humans. Bots seem to be
satisfactorily disseminating opinions and information, although humans
seem to be more attractive (using survey-attractiveness metrics).

\subsection{Comparison to Related Work} \label{ctrw}

We performed a comparison of our detection technique with related work. Specifically, we used the
provided APIs to query the Botometer (formerly BotOrNot) ~\cite{davis2016botornot} and DeBot ~\cite{chavoshi2016debot} services regarding the accounts 
that we found consistently and repeatedly inject similar content respectively for the 3 different collected datasets. 

As Botometer scores range from 0 to 5, we assume that scores over 2,5 are considered bots in this comparison. 
Botometer for the year 2016 provided answers for 907 accounts out of 1052, where 400 out of the 907 had score over 2,5.
For the year 2017 provided answers for 737 accounts out of 947, where 273 out of the 737 had scores over 2,5.
For the year 2018 provided answers for 569 accounts out of 722, where 231 out of the 569 had scores over 2,5.
To summarize, in total we requested for 1850 accounts to be validated and we got answer for the 1414 and the 595 had score over 2,5; 
the rest 436 accounts are either de-activated or suspended accounts.

Debot checks whether the given account has so far been detected, if an account exists in their database, 
it is marked with '1', which means the account has been detected as a bot, otherwise with '0'. 
The date of appearance is also provided (which can be more than one day). 
Due to the daily rate limit imposed by the DeBot service, we requested them to validate all of our 1850 accounts.
Out of the 1850 accounts only 7 were marked as bots, however, this may be an artifact of DeBot restriction to English-speaking accounts, 
or a matter of sample size.

\section{Conclusions}

This paper presents a detection framework for synchronized content
injection on Twitter.  We employed the detection analysis on a set of
720 million tweets and discovered a set of 73 thousand users that
appeared to tweet seemingly independently very similar content.  We
present further analysis of these findings that filter out
coincidental content injection, due to out-of-Twitter reasons and
increase confidence in concluding the existence of content-promoting
botnets. We correlate the detected botnets with crowd-sourced
information mined from Twitter lists and find that the most prolific
botnets are most often targeting specific parts of the Twitter graph,
focusing mainly on news and politics, as well as botnets that promote
web traffic or try to manipulate Twitter trending topics.  We show how
these botnets infiltrate and interact with existing user communities
for many kinds of Twitter interactions.

\section{Acknowledgements}
This document is the results of the research project co-funded by the European Commission, project CONCORDIA, with grant number 830927 (EUROPEAN COMMISSION Directorate-General Communications Networks, Content and Technology) and by the European Union and Greek national funds through the Operational Program Competitiveness, Entrepreneurship and Innovation, under the call RESEARCH – CREATE – INNOVATE (project code:T1EDK-02857, and T1EDK-01800). We would also like to thank the anonymous reviewers for their meaningful and constructive comments that formed the final version of this study.


\begin{thebibliography}{10}
\providecommand{\url}[1]{#1}
\csname url@samestyle\endcsname
\providecommand{\newblock}{\relax}
\providecommand{\bibinfo}[2]{#2}
\providecommand{\BIBentrySTDinterwordspacing}{\spaceskip=0pt\relax}
\providecommand{\BIBentryALTinterwordstretchfactor}{4}
\providecommand{\BIBentryALTinterwordspacing}{\spaceskip=\fontdimen2\font plus
\BIBentryALTinterwordstretchfactor\fontdimen3\font minus
  \fontdimen4\font\relax}
\providecommand{\BIBforeignlanguage}[2]{{%
\expandafter\ifx\csname l@#1\endcsname\relax
\typeout{** WARNING: IEEEtran.bst: No hyphenation pattern has been}%
\typeout{** loaded for the language `#1'. Using the pattern for}%
\typeout{** the default language instead.}%
\else
\language=\csname l@#1\endcsname
\fi
#2}}
\providecommand{\BIBdecl}{\relax}
\BIBdecl

\bibitem{varol2017online}
O.~Varol, E.~Ferrara, C.~A. Davis, F.~Menczer, and A.~Flammini, ``Online
  human-bot interactions: Detection, estimation, and characterization,''
  \emph{arXiv preprint arXiv:1703.03107}, 2017.

\bibitem{subrahmanian2016darpa}
V.~Subrahmanian, A.~Azaria, S.~Durst, V.~Kagan, A.~Galstyan, K.~Lerman, L.~Zhu,
  E.~Ferrara, A.~Flammini, and F.~Menczer, ``The darpa twitter bot challenge,''
  \emph{Computer}, vol.~49, no.~6, pp. 38--46, 2016.

\bibitem{ferrara2016rise}
E.~Ferrara, O.~Varol, C.~Davis, F.~Menczer, and A.~Flammini, ``The rise of
  social bots,'' \emph{Communications of the ACM}, vol.~59, no.~7, pp. 96--104,
  2016.

\bibitem{cao2014uncovering}
Q.~Cao, X.~Yang, J.~Yu, and C.~Palow, ``Uncovering large groups of active
  malicious accounts in online social networks,'' in \emph{Proceedings of the
  2014 ACM SIGSAC Conference on Computer and Communications Security}.\hskip
  1em plus 0.5em minus 0.4em\relax ACM, 2014, pp. 477--488.

\bibitem{twawler}
P.~Pratikakis, ``{twAwler}: A lightweight twitter crawler,'' \emph{arXiv
  preprint arXiv:1804.07748}, 2018.

\bibitem{gephi}
M.~Bastian, S.~Heymann, M.~Jacomy \emph{et~al.}, ``Gephi: an open source
  software for exploring and manipulating networks.'' \emph{Icwsm}, vol.~8, pp.
  361--362, 2009.

\bibitem{ratkiewicz2011detecting}
J.~Ratkiewicz, M.~Conover, M.~R. Meiss, B.~Gon{\c{c}}alves, A.~Flammini, and
  F.~Menczer, ``Detecting and tracking political abuse in social media.''
  \emph{ICWSM}, vol.~11, pp. 297--304, 2011.

\bibitem{igraph}
\BIBentryALTinterwordspacing
G.~Csardi and T.~Nepusz, ``The igraph software package for complex network
  research,'' \emph{InterJournal}, vol. Complex Systems, p. 1695, 2006.
  [Online]. Available: \url{http://igraph.sf.net}
\BIBentrySTDinterwordspacing

\bibitem{strayer2008botnet}
W.~T. Strayer, D.~Lapsely, R.~Walsh, and C.~Livadas, ``Botnet detection based
  on network behavior,'' in \emph{Botnet detection}.\hskip 1em plus 0.5em minus
  0.4em\relax Springer, 2008, pp. 1--24.

\bibitem{thomas2013trafficking}
K.~Thomas, D.~McCoy, C.~Grier, A.~Kolcz, and V.~Paxson, ``Trafficking
  fraudulent accounts: The role of the underground market in twitter spam and
  abuse.'' in \emph{USENIX Security Symposium}, 2013, pp. 195--210.

\bibitem{thomas2010koobface}
K.~Thomas and D.~M. Nicol, ``The koobface botnet and the rise of social
  malware,'' in \emph{Malicious and Unwanted Software (MALWARE), 2010 5th
  International Conference on}.\hskip 1em plus 0.5em minus 0.4em\relax IEEE,
  2010, pp. 63--70.

\bibitem{thomas2011suspended}
K.~Thomas, C.~Grier, D.~Song, and V.~Paxson, ``Suspended accounts in
  retrospect: an analysis of twitter spam,'' in \emph{Proceedings of the 2011
  ACM SIGCOMM conference on Internet measurement conference}.\hskip 1em plus
  0.5em minus 0.4em\relax ACM, 2011, pp. 243--258.

\bibitem{scibot}
H.~Stefanie, D.~B. Timothy, H.~Kim, T.~Andrew, R.~S. Cassidy, and L.~Vincent,
  ``Tweets as impact indicators: Examining the implications of automated
  “bot” accounts on twitter.'' \emph{willey}, 2016.

\bibitem{botwalk}
M.~Amanda, C.~Nikan, K.~Danai, and M.~Abdullah, ``Botwalk: Efficient adaptive
  exploration of twitter bot networks.'' in \emph{Proceedings of the 2017
  IEEE/ACM International Conference on Advances in Social Networks Analysis and
  Mining}, 2017, pp. 467--474.

\bibitem{cyberbot2017}
C.~Zi, G.~Steven, W.~Haining, and J.~Sushil, ``Detecting automation of twitter
  accounts: Are you a human, bot, or cyborg?'' in \emph{IEEE Transactions on
  Dependable and Secure Computing}, 2012, pp. 811 -- 824.

\bibitem{zangerle2014sorry}
E.~Zangerle and G.~Specht, ``Sorry, i was hacked: a classification of
  compromised twitter accounts,'' in \emph{Proceedings of the 29th Annual ACM
  Symposium on Applied Computing}.\hskip 1em plus 0.5em minus 0.4em\relax ACM,
  2014, pp. 587--593.

\bibitem{davis2016botornot}
C.~A. Davis, O.~Varol, E.~Ferrara, A.~Flammini, and F.~Menczer, ``Botornot: A
  system to evaluate social bots,'' in \emph{Proceedings of the 25th
  International Conference Companion on World Wide Web}.\hskip 1em plus 0.5em
  minus 0.4em\relax International World Wide Web Conferences Steering
  Committee, 2016, pp. 273--274.

\bibitem{chavoshi2016debot}
N.~Chavoshi, H.~Hamooni, and A.~Mueen, ``Debot: Twitter bot detection via
  warped correlation.'' in \emph{ICDM}, 2016, pp. 817--822.

\bibitem{freitas2015reverse}
C.~Freitas, F.~Benevenuto, S.~Ghosh, and A.~Veloso, ``Reverse engineering
  socialbot infiltration strategies in twitter,'' in \emph{Proceedings of the
  2015 IEEE/ACM International Conference on Advances in Social Networks
  Analysis and Mining 2015}.\hskip 1em plus 0.5em minus 0.4em\relax ACM, 2015,
  pp. 25--32.

\bibitem{zhang2011detecting}
C.~M. Zhang and V.~Paxson, ``Detecting and analyzing automated activity on
  twitter,'' in \emph{International Conference on Passive and Active Network
  Measurement}.\hskip 1em plus 0.5em minus 0.4em\relax Springer, 2011, pp.
  102--111.

\bibitem{edwards2014bot}
C.~Edwards, A.~Edwards, P.~R. Spence, and A.~K. Shelton, ``Is that a bot
  running the social media feed? testing the differences in perceptions of
  communication quality for a human agent and a bot agent on twitter,''
  \emph{Computers in Human Behavior}, vol.~33, pp. 372--376, 2014.

\end{thebibliography}


\section{Appendix} \label{appendix}
\begin{figure}[h]
\centering
\includegraphics[width=\linewidth]{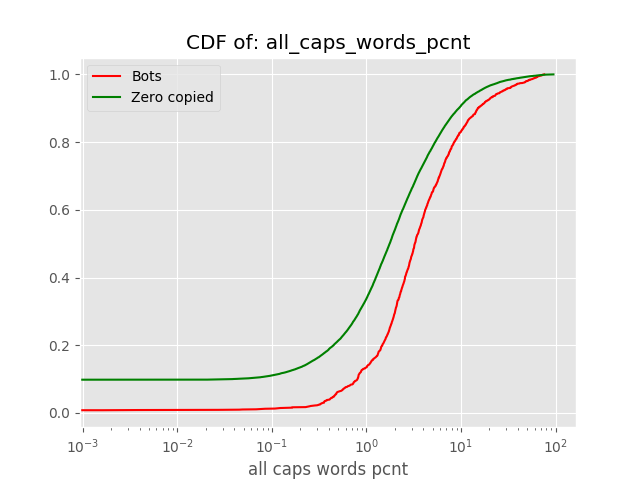}
\caption{CDF distribution of all capital words (percentage of all words) over two different sets of clear users (users with zero copied events in our dataset) and users that were identified as bot (with threshold T of copied events). Figure shows that bot user accounts tend to use less capital words in their tweet text.
}
\label{}
\end{figure}

\begin{figure}[h]
\centering
\includegraphics[width=\linewidth]{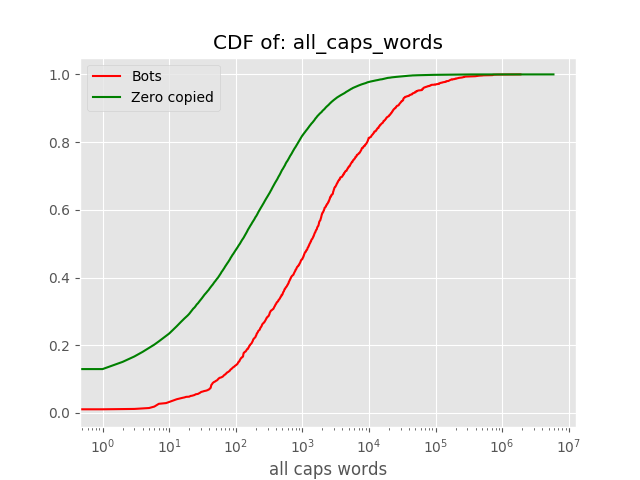}
\caption{CDF distribution of all capital words (count) over two different sets of clear users (users with zero copied events in our dataset) and the users that were identified as bots (with threshold T of copied events). Figure shows that bot user accounts tend to use less capital words in their tweet text.
}
\label{}
\end{figure}

\begin{figure}[h]
\centering
\includegraphics[width=\linewidth]{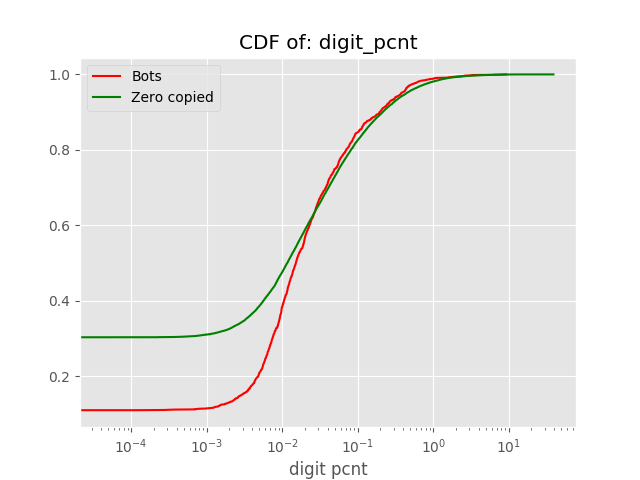}
\caption{CDF distribution of digits/number usage (percentage) over two different sets of clear users (users with zero copied events in our dataset) and the users that were identified as bots (with threshold T of copied events). Figure shows that bot user accounts tend to use less digits/numbers in their tweet text.
}
\label{}
\end{figure}

\begin{figure}[h]
\centering
\includegraphics[width=\linewidth]{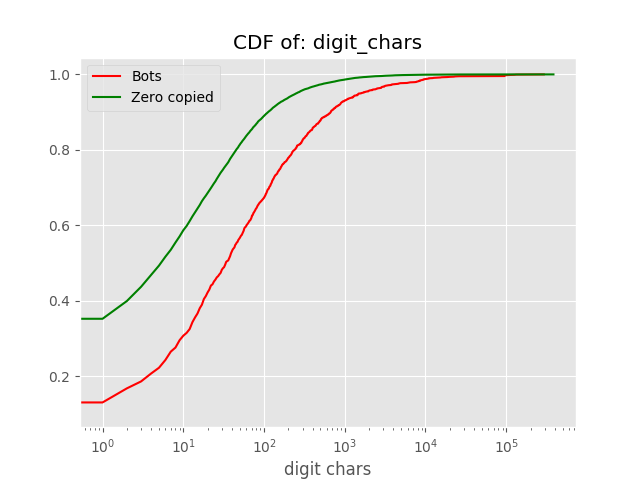}
\caption{CDF distribution of digits/number usage (count) over two different sets of clear users (users with zero copied events in our dataset) and the users that were identified as bots (with threshold T of copied events). Figure shows that bot user accounts tend to use less digits/numbers in their tweet text.
}
\label{}
\end{figure}

\begin{figure}[h]
\centering
\includegraphics[width=\linewidth]{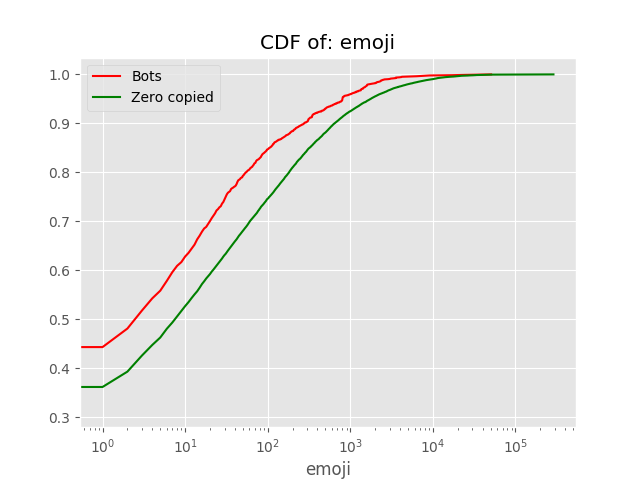}
\caption{CDF distribution of emoji usage (count) over two different sets of clear users (users with zero copied events in our dataset) and the users that were identified as bots (with threshold T of copied events). Figure shows that bot user accounts tend to use more emojis in their tweet text.
}
\label{}
\end{figure}

\begin{figure}[h]
\centering
\includegraphics[width=\linewidth]{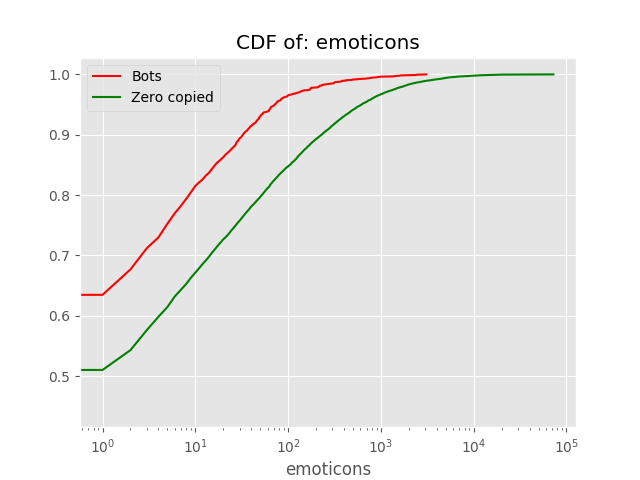}
\caption{CDF distribution of emoticons usage (count) over two different sets of clear users (users with zero copied events in our dataset) and the users that were identified as bots (with threshold T of copied events). Figure shows that bot user accounts tend to use more emojis in their tweet text.
}
\label{}
\end{figure}

\begin{figure}[h]
\centering
\includegraphics[width=\linewidth]{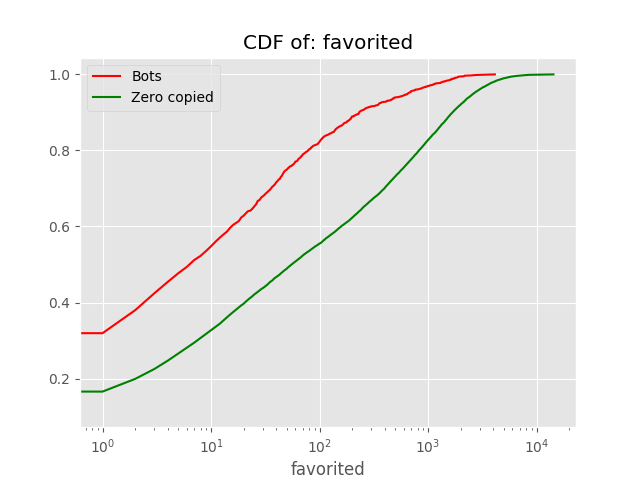}
\caption{CDF distribution of number of favorited accounts (percentage) over two different sets of clear users (users with zero copied events in our dataset) and users that were identified as bots (with threshold T of copied events). Figure shows that bot user accounts tend to have more favorited accounts.
}
\label{}
\end{figure}

\begin{figure}[h]
\centering
\includegraphics[width=\linewidth]{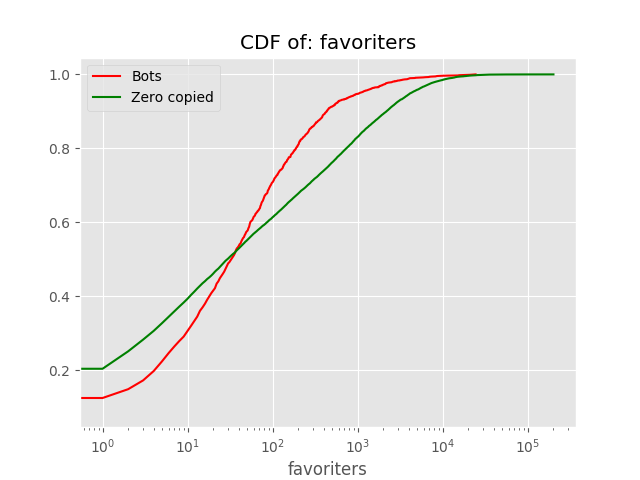}
\caption{CDF distribution of number of favoriters accounts (percentage) over two different sets of clear users (users with zero copied events in our dataset) and the users that were identified as bots (with threshold T of copied events).
}
\label{}
\end{figure}

\begin{figure}[h]
\centering
\includegraphics[width=\linewidth]{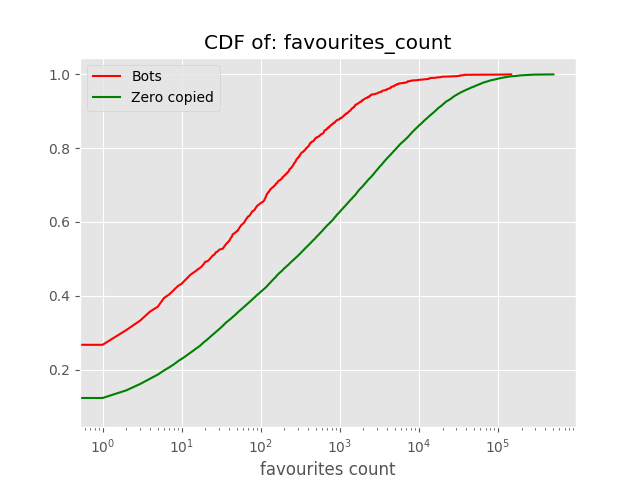}
\caption{CDF distribution of number of favoriters accounts (count) over two different sets of clear users (users with zero copied events in our dataset) and the users that were identified as bots (with threshold T of copied events). Figure shows that bot user accounts tend to have more favoriters accounts.
}
\label{}
\end{figure}

\begin{figure}[h]
\centering
\includegraphics[width=\linewidth]{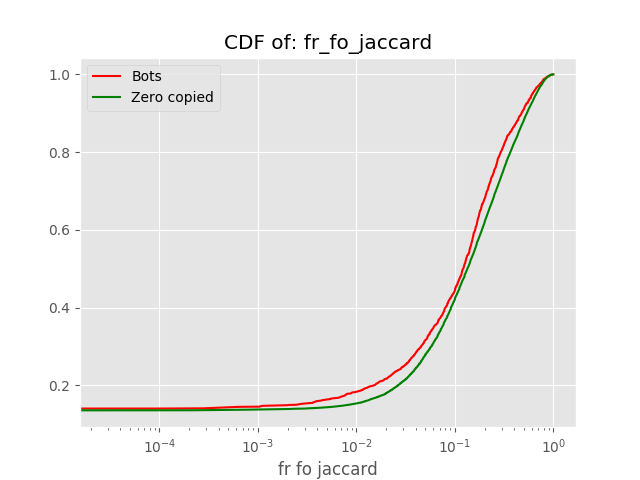}
\caption{CDF distribution of friends to followers Jaccard similarity over two different sets of clear users (users with zero copied events in our dataset) and the users that were identified as bots (with threshold T of copied events). Figure shows that two distributions are very similar.
}
\label{}
\end{figure}

\begin{figure}[h]
\centering
\includegraphics[width=\linewidth]{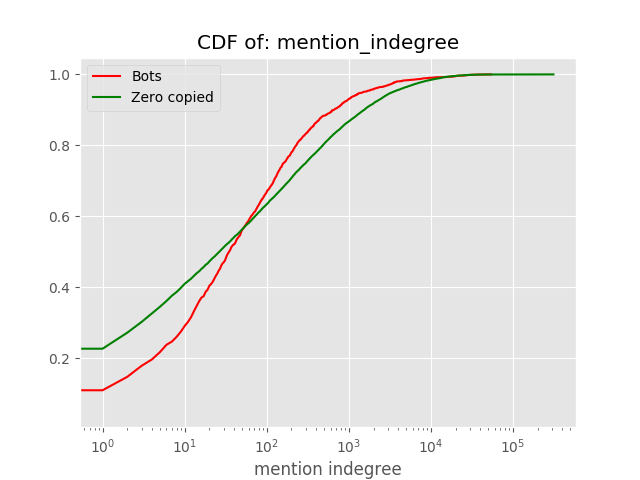}
\caption{CDF distribution of number of mentions of each user (count) over two different sets of clear users (users with zero copied events in our dataset) and the users that were identified as bots (with threshold T of copied events).
}
\label{}
\end{figure}

\begin{figure}[h]
\centering
\includegraphics[width=\linewidth]{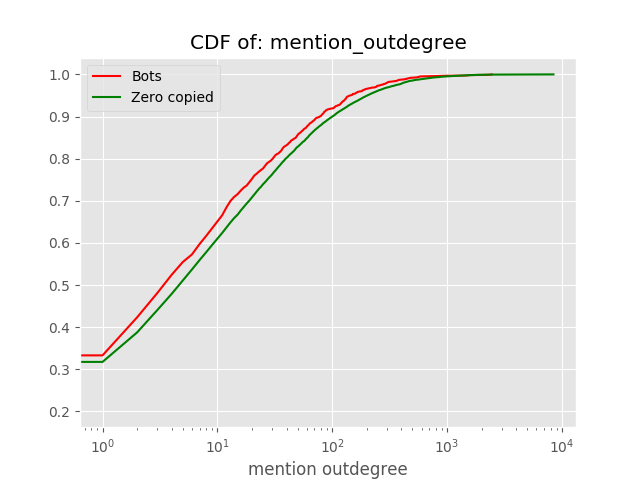}
\caption{CDF distribution of number of mentioned user (count) over two different sets of clear users (users with zero copied events in our dataset) and the users that were identified as bots (with threshold T of copied events).
}
\label{}
\end{figure}

\begin{figure}[h]
\centering
\includegraphics[width=\linewidth]{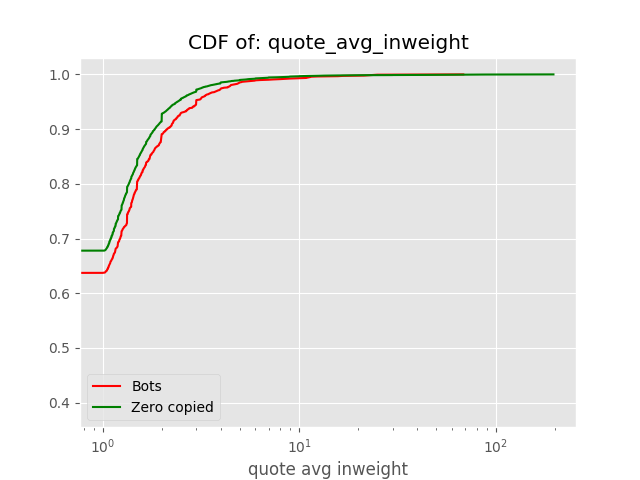}
\caption{CDF distribution of quoted by other users (average with weight) over two different sets of clear users (users with zero copied events in our dataset) and the users that were identified as bots (with threshold T of copied events).
}
\label{}
\end{figure}

\begin{figure}[h]
\centering
\includegraphics[width=\linewidth]{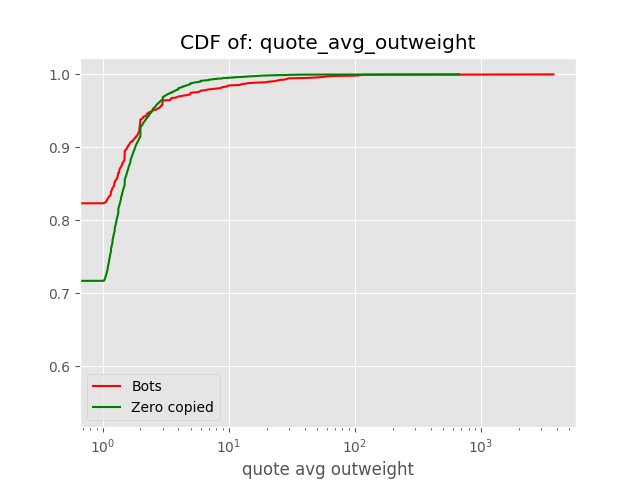}
\caption{CDF distribution of quoted users (average with weight) over two different sets of clear users (users with zero copied events in our dataset) and the users that were identified as bots (with threshold T of copied events).
}
\label{}
\end{figure}

\begin{figure}[h]
\centering
\includegraphics[width=\linewidth]{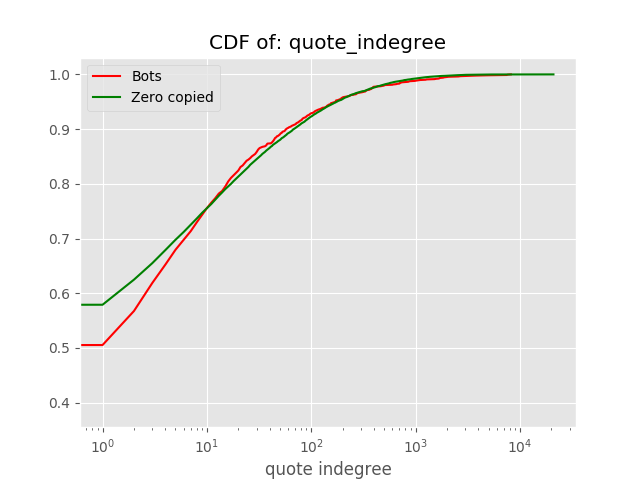}
\caption{CDF distribution of quoted by other users (count degree) over two different sets of clear users (users with zero copied events in our dataset) and the users that were identified as bots (with threshold T of copied events).
}
\label{}
\end{figure}

\begin{figure}[h]
\centering
\includegraphics[width=\linewidth]{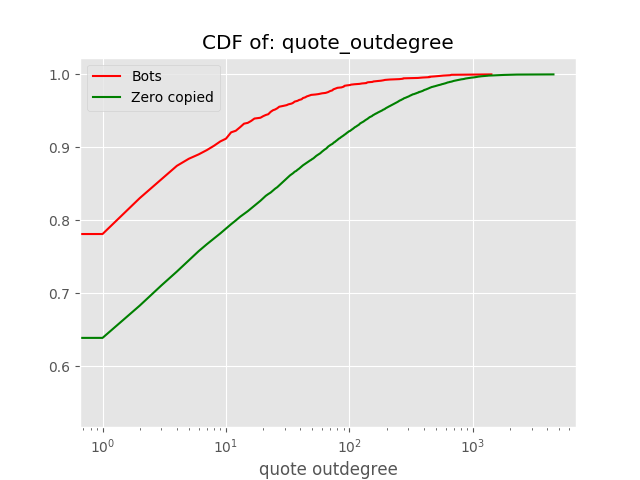}
\caption{CDF distribution of quoted users (count degree) over two different sets of clear users (users with zero copied events in our dataset) and the users that were identified as bots (with threshold T of copied events).
}
\label{}
\end{figure}

\begin{figure}[h]
\centering
\includegraphics[width=\linewidth]{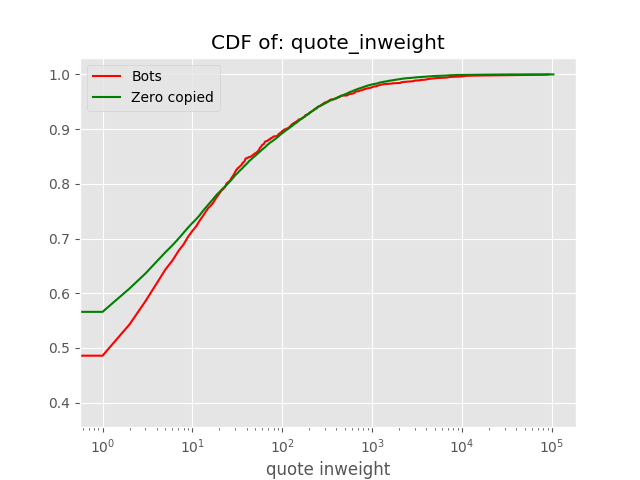}
\caption{CDF distribution of quoted by other users (count weighted) over two different sets of clear users (users with zero copied events in our dataset) and the users that were identified as bots (with threshold T of copied events).
}
\label{}
\end{figure}

\begin{figure}[h]
\centering
\includegraphics[width=\linewidth]{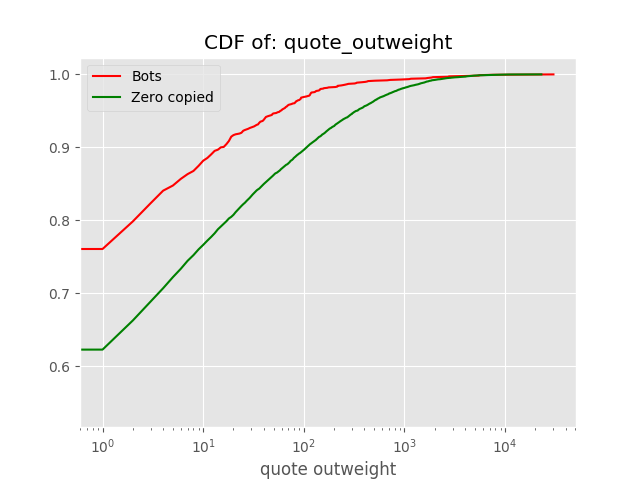}
\caption{CDF distribution of quoted users (count weighted) over two different sets of clear users (users with zero copied events in our dataset) and the users that were identified as bots (with threshold T of copied events).
}
\label{}
\end{figure}

\begin{figure}[h]
\centering
\includegraphics[width=\linewidth]{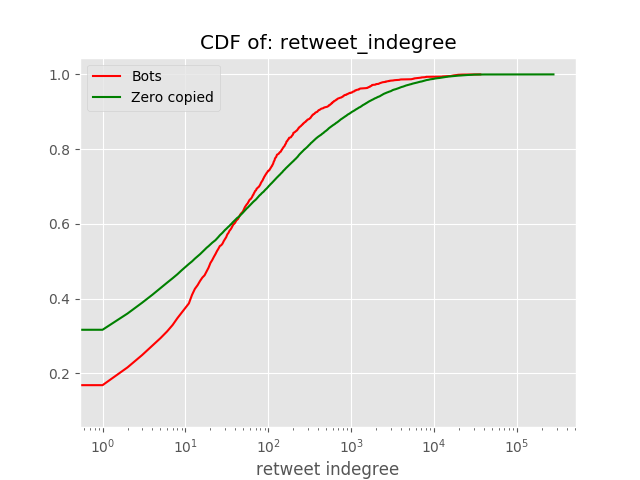}
\caption{CDF distribution of incoming retweets (count degree) over two different sets of clear users (users with zero copied events in our dataset) and the users that were identified as bots (with threshold T of copied events).
}
\label{}
\end{figure}

\begin{figure}[h]
\centering
\includegraphics[width=\linewidth]{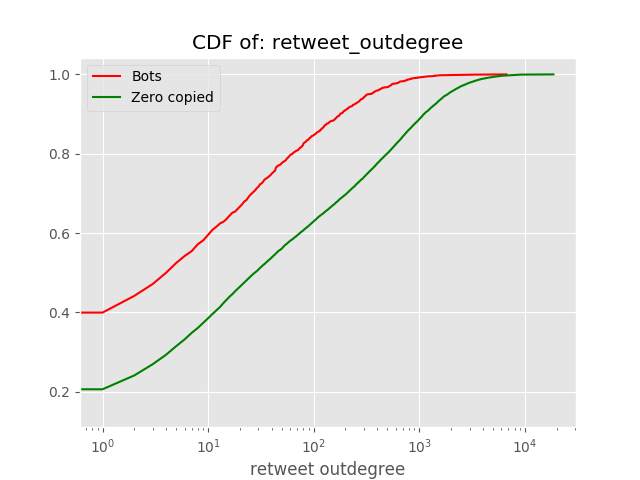}
\caption{CDF distribution of outcoming retweets (count degree) over two different sets of clear users (users with zero copied events in our dataset) and the users that were identified as bots (with threshold T of copied events).
}
\label{}
\end{figure}

\begin{figure}[h]
\centering
\includegraphics[width=\linewidth]{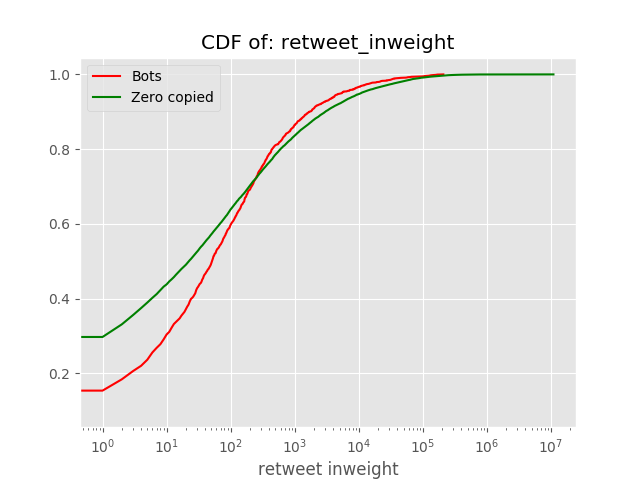}
\caption{CDF distribution of incoming retweets (count weighted) over two different sets of clear users (users with zero copied events in our dataset) and the users that were identified as bots (with threshold T of copied events).
}
\label{}
\end{figure}

\begin{figure}[h]
\centering
\includegraphics[width=\linewidth]{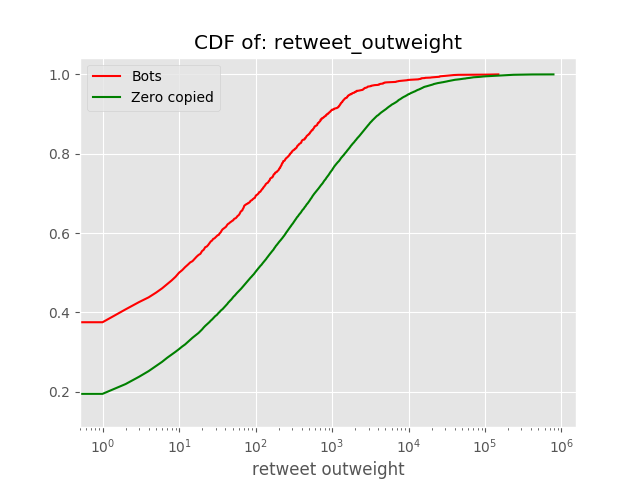}
\caption{CDF distribution of outcoming retweets (count weighted) over two different sets of clear users (users with zero copied events in our dataset) and the users that were identified as bots (with threshold T of copied events).
}
\label{}
\end{figure}

\begin{figure}[h]
\centering
\includegraphics[width=\linewidth]{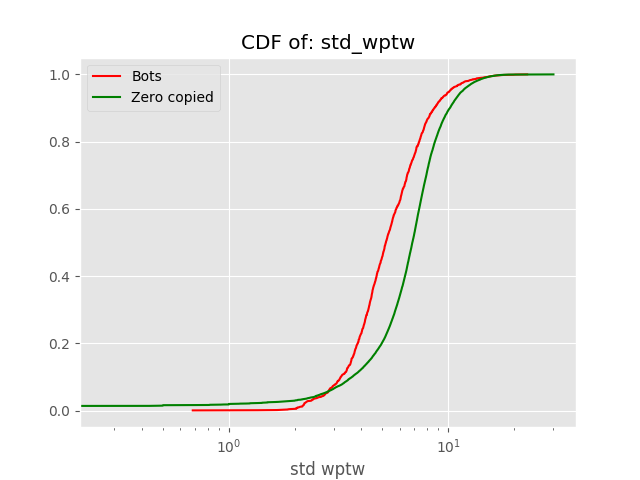}
\caption{CDF distribution of words per tweet (standard deviation) over two different sets of clear users (users with zero copied events in our dataset) and the users that were identified as bots (with threshold T of copied events). User market as bot has a more strict range of words per tweet than usual users.
}
\label{}
\end{figure}

\begin{figure}[h]
\centering
\includegraphics[width=\linewidth]{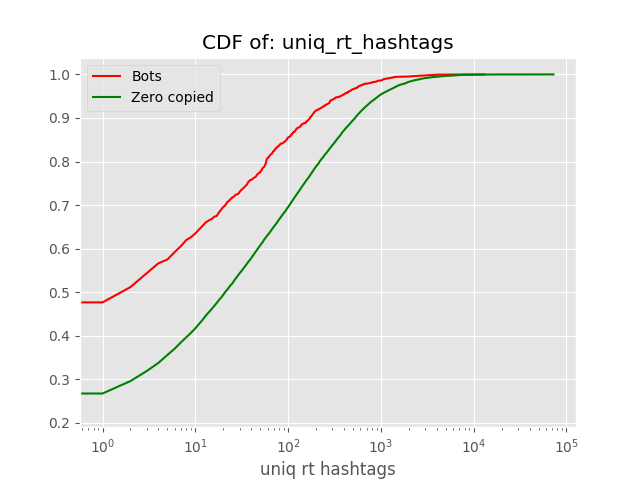}
\caption{CDF distribution of unique retweet hashtags over two different sets of clear users (users with zero copied events in our dataset) and the users that were identified as bots (with threshold T of copied events). In the figure we notice that bot accounts tend to use more unique hashtags in retweeted text.
}
\label{}
\end{figure}

\begin{figure}[h]
\centering
\includegraphics[width=\linewidth]{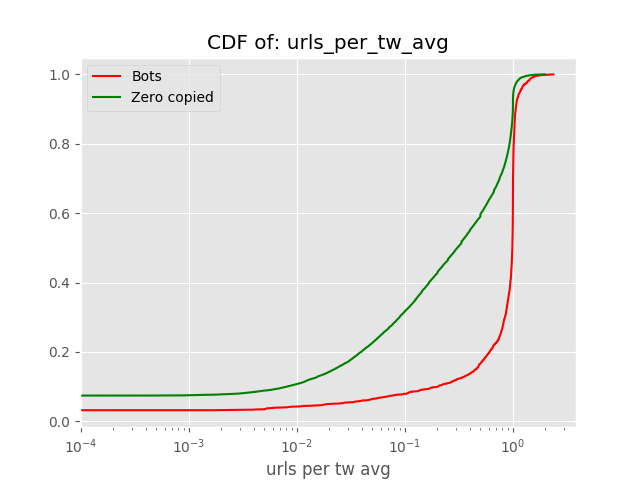}
\caption{CDF distribution of URLs per tweet (average) over two different sets of clear users (users with zero copied events in our dataset) and the users that were identified as bots (with threshold T of copied events). Bot user accounts tend to use less URLs on average in their tweets.
}
\label{}

\end{figure}

\end{document}